%% file: main.tex
\documentclass[review]{elsarticle}
\usepackage{lineno,hyperref}
\modulolinenumbers[5]

\journal{Engineering Applications of Artificial Intelligence}

\usepackage{xcolor}
\usepackage{soul}
\usepackage[utf8]{inputenc}
\usepackage[small]{caption}

\usepackage{multirow}
\usepackage{tikz}
\usepackage{listings}
\usepackage{tikz}
\usetikzlibrary{positioning,arrows,3d,calc,shapes,matrix,fit,shapes.geometric,automata}
\usepackage[nointegrals]{wasysym}
\usepackage{amsmath,tabu,amsfonts,amssymb}
\usepackage{pifont}
\usepackage{subfigure}
\usepackage{multicol}
\usepackage{wrapfig}
\usepackage{lscape}
\usepackage{rotating}
\usepackage{marvosym} 
\usepackage{url}

\newtheorem{proposition}{Property}

\newtheorem{definition}{Definition}
\newtheorem{ex}{Example}

\newtheorem{pf}{Proof}

\newcommand{\indione}{1}
\newcommand{\inditwo}{2}
\newcommand{\indithree}{3}
\newcommand{\indifour}{4}
\newcommand{\actione}{a}
\newcommand{\actitwo}{b}

\newcommand{\trule}{\MVRightarrow~}

\newcommand\ceil[1]{\lceil#1\rceil}
\newcommand{\argmax}{\arg\!\max}
\newcommand\mintinline[2]{\texttt{#2}}

\begin{document}

\begin{frontmatter}

\title{Distributed Algorithm for Matching between Individuals and
  Activities}
\tnotetext[mytitlenote]{This work is part of the PartENS research project
    supported by the Nord-Pas de Calais region (researcher/citizen
    research projects).}

\author{Maxime Morge}
\cortext[mycorrespondingauthor]{Corresponding author}
\ead{maxime.morge@univ-lille.fr}

\author{Antoine Nongaillard}
\address{Univ. Lille, CNRS, Centrale Lille, UMR 9189 - CRIStAL\\
Centre de Recherche en Informatique Signal et Automatique de Lille,
  F-59000 Lille, France}
\ead{antoine.nongaillard@univ-lille.fr}

\begin{abstract}
  In this paper, we introduce an agent-based model for coalition
  formation which is suitable for our usecase. We propose here two
  clearing-houses mechanisms. The first aims at maximizing the global
  satisfaction of the individuals. The second ensures that all
  individuals are assigned as much as possible to a preferred
  activity. Our experiments show that the outcome of our algorithms
  are better than those obtained with the classical
  search/optimization techniques. Moreover, their distribution speeds
  up their runtime.
\end{abstract}

\end{frontmatter}

\section{Introduction}

Multiagent systems (MAS) is a relevant paradigm for the analysis, the
design and the implementation of systems composed of interacting
autonomous entities. In order to design socio-technical systems for
mediation or simulation, MAS allow to model the feedback loops between
heterogeneous actors whose local decision-making leads to complex
global phenomena. One of the challenges of the MAS community is to
facilitate the elicitation of preferences.

Our work is part of a research project that aims to understand and
model feedback dynamics that occur in a group interacting both in
a social network and a real social network.
The usecase we focus here consists of seniors sharing some
activities. This scenario involves thousands of individuals. We aim at
maximizing the activities shared in order to improve the social
cohesion and avoid the isolation of seniors.
We want to propose a social network such that the users form groups to
practice together some activities.  The purpose of this system is to
suggest to each user people with whom to practice these
activities.

In this paper, we present an agent-based model for coalition formation
which is suitable for our usecase. A set of individuals must be
matched together in order to share one of the activities with respect
to the preferences of individuals for their peers and for the
activities. We define decision rules for the elicitation of the
individual and collective preferences with respect to the coalitions
in terms of economic efficiency, fairness, stability and social
cohesion. Based on the bilateral structure of the problem, we propose
here two clearing houses mechanisms which return some matchings where
none activity is oversubscribed. The first one, said "selective", aims
at maximizing the global satisfaction of the individuals. The second
one, said "inclusive", ensures that all individuals are assigned as
much as possible to a preferred activity.

The agent-based modelling allows us to distribute the execution of
these algorithms which are described as agent behaviours. Our
experiments show that the outcomes reached are better than the outcome
reached by mathematical optimization or classical local search
techniques.  Moreover, the decentralization speeds up the runtime (up
to 5 times).

Section~\ref{sec:related} motivates our approach and compares it with
the related works. We introduce our matching problem in
Section~\ref{sec:framework}. We propose two matching algorithms in
Section~\ref{sec:algorithm}. The decentralization of these algorithms
consists of agent behaviours described in
Section~\ref{sec:behaviour}. Section~\ref{sec:exp} and
Section~\ref{sec:application} exhibit our empirical results. Finally,
we conclude with some directions for future work (cf
Section~\ref{sec:discussion}).

\section{Agent-based model of matching}
\label{sec:related}

Social choice theory aims at designing and analyzing collective
decision-making processes which imply a set of agents
selecting or classifying a subset of alternatives among the available
ones.  Contrary to Economy, Computer science is concerned in this
field by the study of algorithms in order to propose operational
methods. We focus here on a particular matching problem.

In order to illustrate our usecase, we present here a walk-through toy
scenario.

\begin{ex}[walk-through scenario]
  Four individuals ($\indione$, $\inditwo$, $\indithree$ and
  $\indifour$) wants to performs passing with one person and six
  juggling props. Having six clubs and six balls, two activities are
  available: the clubs ($\actione$) and the balls ($\actitwo$). Each
  activity can be performed by at most two individuals. According to
  the interests of the individuals for the activities and their
  affinity for their peers (cf Fig.\ref{fig:preference}), all the
  individuals prefer passing clubs rather than balls but not with
  anyone. If $\indione$ and $\inditwo$ like each other, it is not the
  case for all of them. For instance, we will see in this paper that
  $\indithree$ and $\indifour$ like passing each other clubs but they
  prefer staying alone rather than passing balls each other. Beyond
  the combinatorics of such a problem (63 matching in our toy
  example), the search of a harmonious outcome is a complex system.
\end{ex}

\begin{figure}
  \centering
  \begin{minipage}[c]{.26\linewidth}
    \centering
    $
    \begin{tabu}{l|cc}
      & {\bf\actione} & {\bf\actitwo} \\
      \hline
      {\bf v_{\indione}} & .50 & .25 \\
      {\bf v_{\inditwo}} & .50 & .25 \\
      {\bf v_{\indithree}} & .50 & .25 \\
      {\bf v_{\indithree}} & .50 & .25 \\
    \end{tabu}
    $     
  \end{minipage} \hfill
  \begin{minipage}[c]{.66\linewidth}
    \centering
    $      
    \begin{tabu}{l|cccc}
      {\bf j=} & {\bf \indione} & {\bf \inditwo} & {\bf \indithree} & {\bf \indifour}\\
      \hline
      {\bf w_{\indione}(j)} &     & 1.0 & -0.5 & -1.0\\
      {\bf w_{\inditwo}(j)} & 1.0 &     & 0.5  & -1.0\\
      {\bf w_{\indithree}(j)} & 1.0 & 0.5 &      & -1.0\\
      {\bf w_{\indifour}(j)} & 1.0 & 1.0 & -1.0 & \\
    \end{tabu}
    $      
  \end{minipage}
  \vspace{.5cm}
  \vfill
  \begin{minipage}[c]{.1\linewidth}
    \centering
    $
    \begin{tabu}{c}
      \actione \unrhd_{\indione} \actitwo \\
        \actione \unrhd_{\inditwo} \actitwo \\
        \actione \unrhd_{\indithree} \actitwo \\
        \actione \unrhd_{\indifour} \actitwo \\
    \end{tabu}
    $     
  \end{minipage} \hfill
  \begin{minipage}[c]{.82\linewidth}
    \centering
    $
    \begin{tabu}{rcl}
      \{\inditwo,\indione \} \succ_{\indione} &\{\indione \}& \succ_{\indione} \{\indithree,\indione \} \succ_{\indione} \{\indifour,\indione \}\\
      \{\indione , \inditwo\} \succ_{\inditwo} \{\indithree,\inditwo\} \succ_{\inditwo} &\{\inditwo\}& \succ_{\inditwo} \{\indifour, \inditwo\}\\
      \{\indione ,\indithree\} \succ_{\indithree} \{\inditwo,\indithree\} \succ_{\indithree} &\{\indithree\}& \succ_{\indithree} \{\indifour,\indithree\}\\
      \{\indione ,\indifour\} \succ_{\indifour} \{\inditwo,\indifour\} \succ_{\indifour} &\{\indifour\}&  \succ_{\indifour} \{\indithree,\indifour\}
    \end{tabu}
    $     
  \end{minipage}
  \caption{Cardinal preferences (at top) and the deduced ordinal
    preferences (at bottom) of the individuals over the activities (at
    left) and the peers (at right) as stated in Section~\ref{sec:framework}.}%
  \label{fig:preference}
\end{figure}

In the problem of hedonic coalition formation, which has been
formalized in~\cite{drerze80econometrica}, each player is endowed with
a single preference relation over all the coalitions which contain
this player. Our problem is a specialization of the hedonic coalition
problem.  We can represent our problem as a hedonic game. Therefore,
the preferences are not additively separable as in~\cite{aziz13ai},
but it requires the generation of rational lists for coalitions
(RLC)~\cite{balester04geb}. Conversely, the succint representation of
the individual preferences (the preferences over the activities and
over the peers are independent) and the two-sided structure of our
problem, i.e. the fact that the activities are focal
points~\cite{schelling80strategy}, allows us to provide tractable
clearing-houses mechanisms.

The problem of group activity selection has been proposed
in~\cite{darmann12wine}. Each player participates in at most one
activity and its preferences over activities depend on the number of
participants in the activity. This is a generalization of anonymous
hedonic games. This problem has been extended~\cite{igarashi17aaai} in
order to take into account the relationships among the agents, the
latter are encoded by a social network, i.e. an undirected graph where
nodes correspond to agents and edges represent communication links
between them. By contrast, in our problem, each individual is endowed
with a preference relation over the activities and a preference
relation over its peers.

The Hospital/Resident ($HR$) problem has been introduced
in~\cite{gale62ams}. This problem is a specialization of the coalition
formation game where a set of residents must be assigned to the
hospitals in accordance with the preferences of the residents over the
hospitals and the preferences of hospitals over the residents.  The HR
problem has many extensions~\cite{manlove14book}. To the best of our
knowledge, no extension is suitable for our usecase.

How does an agent evaluate its preferences?  In order to reduce the
users' effort for preference elicitation, we adopt here utility
functions (cardinal preferences). Moreover, we assume some additively
separable preferences and that the evaluation of activities and
individuals are comparable. Even if their expressiveness is limited,
the representation of our preferences is linear with respect to the
number of individuals/activities.

What is the ``best'' solution for collective decision-making problem?
In the literature, two kinds of rules derive the social choice from
the individual preferences: the first ones are based on the desirable
properties of the solution (e.g. stability or economic efficiency),
while the latter are based on the aggregation of the individual
preferences (i.e. social welfare), we consider here these two approaches.

How to reach a matching which maximizes the social welfare? It has
been shown in~\cite{everaere13aamas} that DCOP (Distributed Constraint
Optimization Problem) algorithms are not necessarily scalable for
matching problems.  We will see in this paper that mathematical
optimization and local search techniques are not suitable here since
the objective function consists of many local optima.  That is the
reason why we adopt a multiagent approach, in particular a multi-level
model as recommended by~\cite{nongaillard16ecai} where each coalition
agent represents a group of individuals.

\section{Matching between individuals and activities}
\label{sec:framework}

We abstract away from our practical application to introduce a formal
framework for coalition formation which captures our usecase: the
problem, the solutions and their evaluation from an
individual/collective viewpoint.

\subsection{Preferences over individuals and activities}
\label{sec:frameworkProblem}

We introduce here the individuals/activities (IA) problem. In such a
problem, the individuals selects their favorite activities with the
partners they prefer.

\begin{definition}[IA Problem]
  \label{def:ia2}
  An \textbf{individuals/activities (IA) problem} of size $(m,n)$, such that $m
  \geq 2$ and $n \geq 1$, is a couple $IA= \langle I, A \rangle$ with
  $m$ individuals and $n$ activities, where:
  \begin{itemize}
  \item $A=\{a_1, \ldots, a_n\}$ is a set of $n$ activities. Each
    activity $a_j$ has a maximal capacity $c_j \in \mathbb{N}^+$ ;
  \item $I=\{1, \ldots, m\}$ is a set of $m$ individuals. Each
    individual $i$ is endowed with,
    \begin{enumerate}
    \item an interest function over the activities
      $v_i: A \cup \{\theta\} \rightarrow [-1,1]$
    \item an affinity function over the peers
      $w_i: I \setminus \{ i\} \rightarrow [-1,1]$
    \end{enumerate}
  \end{itemize}
\end{definition}
Intuitively, the void activity (denoted~$\theta$), which corresponds
to do nothing, has an infinite capacity : all the individuals can be
assigned to it.

The interest/affinity functions allow to express the repulsion ($<0$),
the attraction ($\geq 0$), and eventually the indifference ($0$) of an
individual with respect to an activity/peer. By convention, the
interest of an individual with respect to the void activity is
neutral:
\begin{equation}
\forall i \in I~v_i(\theta)=0\label{eq:voidactivityinterest}
\end{equation}

We deduce from each interest function $v_i$ an individual preference
relation over the activities, i.e. a reflexive, complete and
transitive preference ordering over $A \cup \{ \theta \}$. This
preference relation (denoted~$\unrhd_i$) is defined such that:
\begin{equation}
  \forall i \in  I ~ \forall a_j, a_k \in A \cup \{ \theta \}~ (v_i(a_j) \geq v_i(a_k) \Rightarrow  a_j \unrhd_i a_k)
  \label{eq:unrhd}
\end{equation}
The corresponding strict relation is denoted~$\rhd_i$.

In order to evaluate the satisfaction of an individual $i$ to belong
to a group $g$, we extend his affinity function over all the groups to
which he belongs, denoted $G(i)=\{g \subseteq I \mid i \in g\}$. Thus,
the satisfaction for a individual to belong to a group only depends on
its peers in this group and not on the other groups formed by the
remaining individuals. In other words, preferences over groups are
purely hedonic preferences. For now, we abstractly define the affinity
of an individual $i$ for a group in the following way:
\begin{align}
  w_i&: G(i) \rightarrow [-1,1] \nonumber\\
  \forall i \in  I ~ \forall g \in G(i)~  w_i(g) &= \circledcirc_{j \in g \setminus \{i\}} ~w_i(j)
\label{eq:wg}
\end{align}
This affinity, which is normalized ($w_i(g) \in [1.1]$), evaluates the
preferences of an individual over $2^{m-1}$ groups. The affinity
function will be instancied in Section~\ref{sec:exp}.

We deduce from each affinity function $w_i$ a preference relation over
the groups of peers, i.e. a reflexive, complete and transitive
preference ordering over $G(i)$. This preference relation
(denoted~$\succsim_i$) is defined such that:
\begin{equation}
  \forall i \in  I ~ \forall g, g' \in G(i)~ (w_i(g) \geq w_i(g') \Rightarrow  g \succsim_i g')
  \label{eq:sucsimm}
\end{equation}
The corresponding strict relation is denoted~$\succ_i$ and the
equivalence one~$\sim_i$.

The satisfaction of an individual $i$ to belong to a group $g$ for
practice an activity $a$ depends on: (i) his or her interest in the
activity; (ii) his affinity with the group. For now, we abstractly
define the utility function in the following manner:
\begin{align}\label{eq:utility}
  u_i: G(i) \times A \cup \{\theta \} \rightarrow [-1,1] \nonumber \\
  \forall i \in I~ \forall g \in G(i) ~ \forall a \in A \cup \{\theta\},~u_i(g,a) = w_i(g) \oplus v_i(a)
\end{align}
We assume that the utility is normalized ($u_i(g,a) \in [-1,1]$).  The
utility function will be instancied in Section~\ref{sec:exp}.

In the rest of the paper, we do not consider that the utilities are
deduced from ordinal preferences as in~\cite{boutilier05ai} but we
suppose a direct access to the utilities of agents since it is the
case in our practical application (e.g. the preferences depicted in
Fig.~\ref{fig:preference}).

\subsection{Matching}
\label{sec:frameworkMatching}

We can now define the possible solutions for the matching problem
previously presented.

We aim at forming coalitions of individuals around the activities.
\begin{definition}[Coalition]
  \label{def:coalition}
  A \textbf{coalition} is a couple $C=\langle a,g \rangle$ where
  $a \in A \cup \{\theta\}$ and $g \subseteq I$. The capacity of the
  coalition (denoted~$c_C$) corresponds to the capacity of the
  activity (denoted~$a_C$) if $a_C \in A$ and $1$ otherwise.  The
  size of the coalition is the cardinality of the group (denoted
  $g_C$). A coalition is said \textbf{sound} if the size does not
  exceed its capability: $\mathrm{card}(g_C) \leq c_C$. A non-empty
  coalition $C$ is such that $g_C\neq \emptyset$. $C$ is for $i$ if
  $i \in g_C$.
\end{definition}

We consider a problem instance such that the number of individuals is
considerably larger than the number of available activities ($m>>n$)
as in our practical application.

\begin{definition}[Matching]
  \label{def:matching}
  A \textbf{matching} $M$ is represented by the functions
  $a_M : I \rightarrow A \cup \{ \theta \}$ and
  $g_M : I \rightarrow \mathcal{P}(I)$ such that:
  \begin{small}
    \begin{align}
      \forall i \in I,~a_M(i) \in A \cup \{ \theta \}\label{eqn:m1}\\
      \forall i \in I ,~i \in g_M(i) \subseteq I\label{eqn:m2}\\
      \forall i \in I ,~a_M(i)= \theta  \Rightarrow g_M(i)=\{i\}\label{eqn:m3}\\
      \forall i \in I~ \forall j \in g_M(i), ~a_M(j)=a_M(i)\label{eqn:m4}\\
      \forall i \in I~ \forall j  \in I \setminus \{i\},~ a_M(i)= a_M(j) \neq \theta \Rightarrow g_M(i)=g_M(j)\label{eqm:5}
    \end{align}
  \end{small}
\end{definition}
The assignment of an individual is an activity, possibly the void
activity (cf equation~\ref{eqn:m1}). In this case, the individual is
said \textbf{inactive}. Each individual is associated with the group
it belongs (cf equation~\ref{eqn:m2}). All the individuals which are
assigned to the void activity are alone (cf
equation~\ref{eqn:m3}). All the individuals which are associated with
each others have the same activity (cf equation~\ref{eqn:m4}) and
reciprocally all the individuals which are assigned to the same
activity, excepted the void activity, are associated with each other
(cf equation~\ref{eqm:5}).

In a matching $M$, the \textbf{participation} function determines the
set of individuals who practice an activity:
\begin{small}
  \begin{align}
    p_M : A \cup \{\theta\} &\rightarrow \mathcal{P}(I) \nonumber\\
     p_M(a) &= \{ i \in I \mid a_M(i)=a \} \label{eq:pM}
  \end{align}
\end{small}
The set of participants for an activity can be empty. If
$a_M(i)=\theta$, $i$ is said inactive. In a matching $M$, an
activity $a \in A \cup \{ \theta\}$ is:
\begin{enumerate}
\item oversubscribed $\mathrm{card}(p_M(a)) > c_a$ ; 
\item undersubscribed $\mathrm{card}(p_M(a)) < c_a$ ; 
\item full (denoted~$\mathrm{full}_M(a)$) otherwise.
\end{enumerate}
A matching is said \textbf{sound} if no activity is oversubscribed.

A matching $M$ is a coalition structure, i.e. a partition of individuals:
\begin{eqnarray}   
  \forall a \in ~ A \cup \{\theta\},~  \langle a, p_M(a) \rangle \mbox{ is a coalition} \label{eq:partition1}\\
  \bigcup_{a \in A \cup \{ \theta \}} p_M(a)= I \label{eq:partition2}\\
  \forall a_i \in A \cup \{ \theta \}~\forall a_j \in A \cup \{ \theta \} \setminus \{a_i\}, \nonumber\\
  p_M(a_i) \cap p_M(a_j) = \emptyset \label{eq:partition3}
\end{eqnarray}
Indeed, the equations~\ref{eq:partition1}, \ref{eq:partition2}
and~\ref{eq:partition3} are deduced from the equation~\ref{eq:pM}, the
definition~\ref{def:coalition} and the definition~\ref{def:matching}.
We denote $C_M(i)$ the coalition in $M$ which contains $i$.  In order
to simplify the definitions about stability in the next section, we
introduce the \textbf{target} function which returns the newcomers by
selecting the activity $a$ in the matching $M$:
\begin{align}
  o_M : A \cup \{\theta\} &\rightarrow \mathcal{P}(i) \nonumber\\
  o_M(a) =&\begin{cases}
    p_M(a) & \text{ if } a \in A,\\
    \emptyset & \text{otherwise}.
  \end{cases}
\end{align}

\subsection{Individual rationality}
\label{sec:frameworkRationality}

We can now evaluate the satisfaction of the individuals with respect
to the possible solutions.

Each individual assesses the coalition to which he belongs and thus
the matching based on his group and his activity. An individual $i$
considers that a sound coalition $C$ for him ($i \in g_C$) is
individually rational if and only if $u_{i}(g_C, a_C) \geq 0$.
A sound matching $M$ is \textbf{individually rational}
(denoted~IR) if and only if:
\begin{equation}
  \forall i \in I,~ u_{i}(g_M(i),a_M(i)) \geq 0 
\label{eq:Mrat}
\end{equation}
The existence of a sound IR matching is guaranteed, e.g. the trivial
matching (denoted $M_0$) where all individuals are inactive.

We deduce from each utility function $u_i$ an individual preference
relation over the coalitions, i.e. a reflexive, complete and
transitive preference ordering over $A \cup \{ \theta \} \times G(I)$.
The individual $i$ prefers $C$ rather than $C'$
(denoted~$C \succsim_i C'$) if
$u_{i}(g_{C},a_{C}) \geq u_{i}(g_{C'},a_{C'})$. An individual prefers
a matching to another one if he prefers his coalition in the first.
Let us consider $M$ and $M'$ two matchings for our problem instance
$IA$. The individual $i$ prefers $M$ rather than $M'$ (denoted
$M \succsim_i M'$) if and only if
\begin{equation}
C_M(i) \succsim_i C_{M'}(i) \label{eq:Mpref}
\end{equation}
The strict preference relation over the matchings/coalitions is 
denoted $\succ_i$.

\begin{ex}[Rationality]
\label{ex:matching}
Let us consider the $IA$ problem depicted in Fig.~\ref{fig:preference},
we focus on $3$ of $63$ possible sound matchings:
\begin{itemize}
\item $M_0$ with $p_{M_0}(\actione)= \emptyset$,
  $p_{M_0}(\actitwo)=\emptyset$ and $p_{M_0}(\theta)=\{1, 2, 3, 4\}$;
\item $M_1$ with $p_{M_1}(\actione)=\{ \indione, \inditwo\}$,
  $p_{M_1}(\actitwo)=\{ \indifour\}$ and $p_{M_1}(\theta)=\{ \indithree\}$;
\item $M_2$ with $p_{M_2}(\actione)=\{ \indione, \inditwo\}$,
  $p_{M_2}(\actitwo)=\{ \indithree, \indifour\}$ and $p_{M_2}(\theta)=\emptyset$;
\end{itemize}
Contrary to $M_0$ and $M_1$, $M_2$ is not individually rational since
the sound matching
$\langle \actitwo, \{ \indithree, \indifour \}\rangle$ is neither
rational for $\indithree$, nor for $\indifour$. These two individuals
strictly prefer being inactve rather than being poorly accompanied for
$\actitwo$. By contrast, they strictly prefer practice $\actione$
together rather than being inactive:
\begin{eqnarray}
\langle \actione , \{\indithree, \indifour\} \rangle
\succ_{\indithree} \langle \theta , \{\indithree\} \rangle \succ_{\indithree} \langle
\actitwo , \{\indithree, \indifour\} \rangle \label{eq:exRat1} \\
\langle \actione , \{\indithree, \indifour\} \rangle
\succ_{\indifour} \langle \theta , \{\indifour\} \rangle \succ_{\indifour} \langle
\actitwo , \{\indithree, \indifour\} \rangle \label{eq:exRat2}
\end{eqnarray}
\end{ex}

\subsection{Social evaluation}
\label{sec:frameworkSocialEvaluation}

We can now evaluate the matchings from the collective viewpoint.

A first desirable property for matchings is stability. Intuitively,
the individuals in a blocking coalition would like to separate and
form their own coalition, which makes the underlying matching
unstable.

A non-empty sound coalition $C$ \textbf{strongly blocks} a matching
$M$ if and only if all the individuals of the coalition strictly
prefer the latter to be assigned according to $M$:
\begin{equation}\label{eqn:sblock}
  \forall i \in g_C,~ C \succ_i C_M(i) 
\end{equation}
A sound matching is \textbf{core stable} (CS) if and only if no
non-empty sound coalition which strongly blocks it. By definition, a
matching which is core stable is individually rational.

A non-empty sound coalition $C$ \textbf{weakly blocks} a
matching $M$ if and only if  all the individuals of the coalition
prefer the latter to be assigned according to $M$ and at least 
one individuals strictly prefers the coalition:
\begin{equation}\label{eqn:block}
  \forall i \in g_C,~ C \succsim_i C_M(i) \wedge \exists i \in g_C,~ C \succ_i C_M(i) 
\end{equation}
A sound matching is \textbf{strict core stable} (SCS) if no non-empty
sound coalition strongly blocks it.  By definition, all the SCS
matching are \textit{a fortiori} CS.

A sound matching $M$ est \textbf{Nash stable} (NS) if and only if no
individual can benefit by moving from its coalition toward another
undersubscribed activity:
\begin{equation}\label{eq:ns}
  \begin{aligned}[t]
    &\forall i \in I~\forall a \in A \cup \{\theta\} \setminus \{ a_M(i) \}\\
    &\mathrm{full}_M(a) \vee C_M(i) \succsim_i \langle a,o_M(a) \cup \{ i \} \rangle
  \end{aligned}
\end{equation}
Such a matching is immune to individual movements since the coalitions
are rational for all the individuals and at least as good as the other
coalitions which can be reached by a single move.

A sound matching $M$ is \textbf{individually stable} (IS) if and only
if no individual can benefit by moving from his coalition to another
undersubscribed activity or if so, the deviating individual is not
unanimously appreciated by his newcomers partners:
\begin{equation}\label{eq:is}
  \begin{aligned}[t]
    &\forall i \in I~\forall a \in A \cup \{a\} \setminus \{a_M(i) \}\\
    &\mathrm{full}_M(a)\\
    \vee &\big[ \langle a,o_M(a) \cup \{ i\}\rangle \succ_i
    C_M(i) \Rightarrow\\ 
    &\exists j \in o_M(a),~ C_M(j) \succ_j \langle
    a,o_M(a) \cup \{ i\} \rangle \big]
  \end{aligned}
\end{equation}
By definition, a matching IS is \textit{a fortiori} NS.

A sound matching $M$ is \textbf{contractually individually stable}
(CIS) if and only if no individual can benefit by moving from his
coalition to another without decreasing the satisfaction of one side or
the other:
\begin{equation}\label{eq:cis}
  \begin{small}
    \begin{aligned}[t]
      &\forall i \in I~\forall a \in A \cup \{a\} \setminus \{a_M(i)\}\\
      &\mathrm{full}_M(a)\\
      \vee &\big[ \langle a,o_M(a) \cup \{ i\} \rangle \succ_i
      C_M(i) \Rightarrow\\
      &\exists j \in o_M(a),~ C_M(j) \succ_j \langle  a,o_M(a) \cup \{ i\} \rangle\\
      &\vee \exists j' \in g_M(i)\setminus \{i \}~, C_M(i) \succ_{j'}
      \langle a_M(i), g_M(i) \setminus \{ i \} \rangle \big]
    \end{aligned}
  \end{small}
\end{equation}
By definition, a CIS matching is \textit{a fortiori} IS. It worth
noticing that, contrary to a IS matching, a CIS matching is not
necessarily IR.

We are now defining other desirable properties to evaluate the
economic efficiency of a matching. Like the previous ones, theses
properties are not based on inter-personal comparisons.

A sound $M$ is \textbf{perfect} (P) if the coalition for each individual
is one of his preferred one:
\begin{equation}
  \label{eq:perfect}
  \begin{split}
    &\forall i \in I~ \forall a \in A \cup \{\theta\}~ \forall g
    \in G(i),~ C_M(i) \succsim_i \langle a, g \rangle     
  \end{split}
\end{equation}

A matching $M'$ \textbf{Pareto-dominates} $M$ if and only if $M'$ is strictly
preferred to $M$ for at least one individual and not worst for the
others:
\begin{eqnarray}  
  &\exists i \in I, C_{M'}(i) \succ_i C_M(i) \label{eqn:p1}\\
  \wedge&\forall i \in I, C_{M'}(i) \succsim_i C_M(i) \label{eqn:p2}
\end{eqnarray}
A matching is \textbf{Pareto-optimal} (PO) if it is not
Pareto-dominated.  In other words, A matching is PO if there is no
alternative in which all agents would be in an equivalent or better
position.

The strict core stability is a sufficient condition for the Pareto-optimality.

\begin{proposition}[Efficiency vs. Stability] 
  $\mathrm{SCS} \subseteq \mathrm{PO}$
 \end{proposition}

\begin{pf}[Efficiency vs. Stability] 
  We prove by contradiction that a sound matching which is strict core
  stable is also Pareto-optimal.  We assume that $M$ is a strict core
  stable matching which is not a Pareto-optimum.  Then, there exists a
  matching $M'$ which Pareto-dominates $M$. Let us consider the
  coalition $C_{M'}(i)$ with $i$ satisfying the equations~\ref{eqn:p2}
  and \ref{eqn:p2}. The equation~\ref{eqn:block} is true. Therefore,
  we conclude that $C_{M'}(i)$ is sound and weakly blocks $M$. This is
  in contradiction with our assumption.
\end{pf}

Contrary to the Paretian approach, the social choice theory analyses
the interpersonal comparisons~\cite{sen70socialchoice}. This theory
reconciles the utilitarianism of Jeremy Bentham and the distributive
justice advocated by John Rawls~\cite{moulin02fair}.

\begin{definition}[Social welfare]
  \label{def:sw}
  Let $M$ be a sound matching.
  \begin{itemize}
  \item The \textbf{utilitarian welfare} of $M$ is:
    \begin{equation}\label{eq:swu}
      U(M)=\frac{1}{m}\sum_{i \in I}
      u_{i}(g_M(i),a_M(i))
    \end{equation}
  \item The \textbf{egalitarian welfare} of $M$ is:
    \begin{equation}\label{eq:swe}
      E(M)=\min_{i \in I} (u_{i}(g_M(i),a_M(i)))
    \end{equation}
  \end{itemize}
\end{definition}
The welfares are normalized ($U(M), E(M) \in [-1,1]$). The higher the
welfare is, the better the matching is. A \textbf{maximum utilitarian}
(respectively \textbf{maximum egalitarian}) - denoted MaxUtil
(respectively MaxEgal)- matching maximizes the utilitarian
(resp. egalitarian) welfare. It is worth noticing that, even all the
individuals/activities are attractive, the egalitarian welfare is $0$
if at least one individual is inactive.  By definition, a sound
MaxUtil matching is \textit{a fortiori} PO. Similarly, a sound MaxUtil
matching is \textit{a fortiori} IR.

Finally, Social Psychology points out that cohesion is more closely
linked to the interest of the activities rather than the interpersonal
affinities~\cite{zaccaro88jsp}. Formally, a sound matching $M$ is
\textbf{socially cohesive} (SC) if and only if all individuals are
assigned as much as possible to the activities they prefer:
\begin{equation}
  \label{eq:sc}
  \begin{aligned}[t]
    &\forall i \in I\\
    &\forall a \in \{a \in A \setminus \{ a_M(i)\} \mid v_i(a) \geq 0 \wedge v_i(a) > v_i(a_M(i)) \}\\
    &\mathrm{full}_M(a)
  \end{aligned}
\end{equation}
The existence of a sound SC matching is guaranteed, e.g. the trivial
matching when all activities are rebutting. Inspired by our practical
application which aims at maximizing the activities shared (cf
Section~\ref{sec:application}), this notion is orthogonal to the
previous ones.

\begin{figure}
  \begin{center}
    \begin{tikzpicture}[->, node distance=50pt,
      edge/.style={style=''-latex''}]
      \node (perfect) {P$^{(0)}$}; 
      \node (sc) [below of=perfect] {SCS$^{(0)}$};
      \node (ns) [left of=sc] {NS$^{(7)}$}; 
      \node (maxegal) [left of=ns] {\underline{MaxEgal}$^{(2)}$}; 
      \node (c) [left of=maxegal] {\underline{SC}$^{(6)}$}; 
      \node (maxutil) [right of=sc] {\underline{MaxUtil}$^{(2)}$}; 
      \node (is) [below of=ns] {IS$^{(9)}$}; 
      \node (c) [below of=sc] {CS$^{(0)}$}; 
      \node (po) [below of=maxutil] {\underline{PO}$^{(15)}$}; 
      \node (cis) [below of=po] {\underline{CIS}$^{(16)}$};
      \node (ir) [below of=is] {\underline{IR}$^{(51)}$};  
      \path 
        (perfect) edge (sc) 
        (perfect) edge (ns) 
        (perfect) edge (maxegal) 
        (perfect) edge (maxutil) 
        (maxegal) edge (ir)
        (ns) edge (is) 
        (sc) edge (c) 
        (sc) edge (po)
        (maxutil) edge (po)
        (is) edge (ir)
        (is) edge (cis)
        (c) edge (ir)
        (po) edge (cis);
     \end{tikzpicture} 
   \end{center}
   \caption{Inclusive relationships between stability, optimality,
     fairness and social cohesion.  The existence of the underlined
     concepts is guaranteed. The superscripts is the distribution of
     teh $63$ sound matching for our example (see
     figure~\ref{fig:preference}) depending on their properties.}
   \label{fig:stability}
 \end{figure}
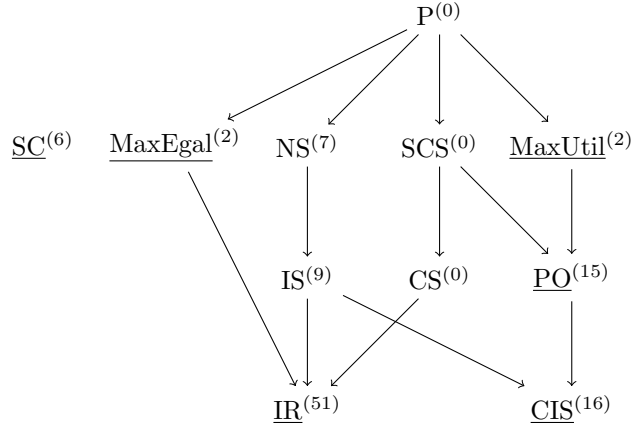

Inspired by~\cite{aziz13ai}, Fig.~\ref{fig:stability} depicts the
inclusion relationships between the previous concepts as a
lattice. For instance, the arc between SCS and PO indicates that the
first concept is a necessary condition for the second but not a
sufficient one as illustrated in the following example.

\begin{ex}[Social evaluation]
Let us consider the $IA$ problem represented in Fig.~\ref{fig:preference},
the sound MaxtUtil matching are :
\begin{itemize}
\item $M_1$ with $p_{M_1}(\actione)=\{ \indione, \inditwo\}$,
  $p_{M_1}(\actitwo)=\{ \indifour\}$ and $p_{M_1}(\theta)=\{ \indithree\}$;
\item $M'_1$ with $p_{M'_1}(\actione)=\{ \indione, \inditwo\}$,
  $p_{M'_1}(\actitwo)=\{ \indithree \}$ and $p_{M'_1}(\theta)=\{ \indifour \}$.
\end{itemize}
The sound MaxEgal matching are:
\begin{itemize}
\item $M_3$ with $p_{M_3}(\actione)=\{ \indione, \indifour\}$,
  $p_{M_3}(\actitwo)=\{ \inditwo, \indithree\}$ and $p_{M_3}(\theta)=\emptyset$;
\item $M'_3$ with $p_{M'_3}(\actione)=\{ \indithree, \indifour\}$,
  $p_{M'_3}(\actitwo)=\{ \indione, \inditwo\}$ and $p_{M'_3}(\theta)=\emptyset$.
\end{itemize}
The matchings $M_1$ and $M'_1$ are PO but not SCS. Contrary to $M_1$
and $M'_1$, the matchings $M_3$ and $M'_3$ are SC but not IR.
\end{ex}

In summary, if the stability is a desirable property, there is no
necessarily such a solution. Moreover, Pareto-optimality doesn't seem
to be discriminating enough. By contrast, the concepts of social
cohesion and social welfare are better candidate to assess the quality
of a matching from the collective viewpoint.

\section{Centralized matching mechanism}
\label{sec:algorithm}

Based on the bilateral structure of the IA problem, we propose here
two algorithms which return sound matchings.

Our algorithms are inspired by the student-proposing deferred
acceptance algorithm~\cite{gale62ams}. Each individual $i$ is endowed
with a list of concessions, i.e. rational activities in decreasing order of
preference. On one side, the individual $i$ proposes itself to the
activities he prefers, even if it must concede in case of
rejection. On the other side, the coalition $\langle a, g \rangle$, as
a host, apply its capacity constraint $c_a$ and the objective of the
hosted individuals, i.e. a social decision rule which can be:
\begin{itemize}
\item either a utilitarian rule 
  \begin{equation}\label{eq:utilitarianSocialRule}
    \argmax_g{\{\sum_{i \in g} u_i(g,a) \mid card(g) \leq c_a \}},
  \end{equation}
\item or an egalitarian rule
  \begin{equation}\label{eq:egalitarianSocialRule}
    \argmax_g{\{\min_{i \in g} u_i(g,a) \mid card(g)\ \leq c_a\}}.    
\end{equation}
\end{itemize}

The initial matching is the trivial one, i. e. all individuals are
inactive. Each in turn, a free individual $i$ considers this prefers
activity which attracts him $a$.  If the coalition is empty, $i$ is
assigned. Otherwise, the algorithms try to improve the satisfaction of
this coalition, eventually by excluding the individuals whose presence
penalizes the (utility or egalitarian) rule of the group. The
individuals who are excluded, eventually $i$, must concede and so
consider the next attractive activity.  

In the \textbf{selective algorithm},the social decision rule is
evaluated for each proposal to a non-emtpy group. As long as the
capacity of $a$ is not reached then the group can grow. Finally, an
individual who is rejected by all his attractive activities is
definitely inactive. Since a group of $k$ individuals contains $2^k-1$
non-emtpy subgroups, we propose an approximation algorithm which
excludes no more than one individual at each step.


Our (approximation) algorithm always returns a sound matching.

\begin{proposition}[Termination]
   \label{prop:termination}
   Our (approximation) algorithm applied over an IA problem instance
   ends and returns a sound matching.
 \end{proposition}

\begin{pf}[Termination]
  \label{proof:termination}
  We consider the following loop invariant:
  $\Sigma_{i \in I} |concessions(i)| + |free|$ with the list of the
  rational activities for $i$: $concessions(i)$ and the set of free
  individuals $free$. This invariant is positive. It strictly
  decreases after each loop since:
\begin{enumerate}
\item an individual, which is assigned, is removed from $Free$;
\item an individual, which is not assigned, concedes until it is
  definitely assigned to the void activity;
\item any individual, which is unassigned, concedes and at least one
  another individual is assigned (and so removed from $free$).
\end{enumerate}
The resulting matching is sound since the activities are never
oversubscribed.
\end{pf}

\begin{ex}[Selective algorithm.]
 \label{ex:algoselectif}
 Let us consider the $IA$ problem represented in
 Fig.~\ref{fig:preference}. Whatever the social decision rule is, our
 selective algorithm allows to reach $M_1$ with
 $p_{M_1}(\actione)=\{ \indione, \inditwo\}$,
 $p_{M_1}(\actitwo)=\{ \indifour\}$ and
 $p_{M_1}(\theta)=\{ \indithree\}$. This matching is Pareto-optimal
 but it is not socially cohesive. 
\end{ex}


It is worth noticing that the selective algorithm does not not
necessarily maximizes the utilitarian (respectively egalitarian)
welfare by applying the utilitarian (respectively egalitarian) rule
since the individuals give priority to the activities over the
individuals. Despite this priority, the resulting matching is not
necessarily socially cohesive. That is the reason why we propose the
inclusive algorithm. In the latter, the social decision rule of a
coalition, as a host, is assessed only if its capacity constraint is
violated and the exclusion of a single individual is
envisaged for each individual integration.


The inclusive algorithm always returns a sound match that is socially
cohesive whatever the social decision rule is.

\begin{proposition}[Termination]
   \label{prop:termination}
   The inclusive algorithm applied over an IA problem instance ends
   and returns a sound matching.
 \end{proposition}

\begin{pf}[Terminaison]
  \label{proof:termination2}
  See proof~\ref{proof:termination}.
\end{pf}

\begin{proposition}[Social cohesion]
   \label{prop:cohesion}
   The inclusive algorithm applied over an IA problem instance returns
   a SC matching.
 \end{proposition}

\begin{pf}[Social cohesion]
  \label{proof:cohesion}
  We prove by contradiction that the outcome $M$ of the inclusive
  algorithme is a SC matching.  We assume that ther exists an
  individual $i$ and an activity $a \neq a_M(i)$ such that
  $v_i(a) \geq 0$, $v_i(a) > v(a_M(i))$ and $a$ is undersubscribed (cf
  equation~\ref{eq:sc}). Since $i$ is attracted by $a$, this
  individual proposes himself to this activity which is in his list of
  concessions (line~4) before $a$. He has been excluded by this
  activity which is full since the size of the possible groups is
  $c_a$ (line~22). This is in contradiction with our assumption.
\end{pf}

\begin{ex}[Inclusive algorithm]
 \label{ex:algoselect}
 Let us consider the $IA$ problem instance depicts in
 Fig.~\ref{fig:preference}. whatever the social decision rule is, the
 inclusive algorithm allows to reach $M_2$ with
 $p_{M_2}(\actione)=\{ \indione, \inditwo\}$,
 $p_{M_2}(\actitwo)=\{ \indithree, \indifour\}$ and
 $p_{M_2}(\theta)=\emptyset$. This matching is socially cohesive but
 it is not individually rational. 
\end{ex}

\section{Agent behaviours}
\label{sec:behaviour}

We describe here the previous matching mechanisms as agent behaviours
in order to decentralize their execution.

We consider the asynchronous message-passing model of actor for
concurrent programming introduced in~\cite{clinger81foundations}.  In
this framework, the primitives are agents and events: an agent is an
independent program that runs on its own processor; an event
corresponds to the creation of an agent or the utterance/reception of
a message. The system is distributed since the underlying channels are
assumed to be reliable (a message is delivered once and only once) and
that the messages may arrive in different order from sending.

In order to propose a distributed solver based on this model, we
distinguish 3 kinds of agents:
\begin{enumerate}
\item the solver agent which creates the other agents and records
  the assignments;
\item the individual agents which are endowed with the same
  behaviours but with different preferences;
\item the activity agents which are endowed with the same behaviours
  but with different capacity constraints and group objectives.
\end{enumerate}

The behaviour of the solver agent consists of: i) creating other
agents; ii) triggering the solving; iii) recording the
assignments/unassignments; and iv) returning the matching when all the
individual are assigned or definitely inactive.

Whether the mechanism is selective or inclusive, the behaviour of an
individual agent is the same. An individual agent proposes himself
to 
his preferred activity that attracts him. When the agent is assigned
or excluded, he informs the solver agent. For an unassignment, the
coalition agent waits for a confirmation before sending a new proposal
such that a matching is not prematurely returned by the solver
agent. If an individual agent becomes free, he concedes, then he
proposes himself to the next attractive activity until being
definitely inactive. In accordance with the principle of subsidiarity,
the reconfiguration of a coalition may be subject to arbitration by
the the coalition agent that requires the opinion of the individual
agents which are concerned.


In order to describe the coalition agent behaviour, we consider 
the finite state machine design pattern for concurrent programming
with 3 states:
\begin{enumerate}
\item the inital \texttt{Disposing} state for proposal processing;
\item the \texttt{Casting} state where the opinion of individual
  agents are requested in order to assess the group in conformance
  with the social decision rule;
\item the \texttt{Firing} state where the excluded individual agents
  are notified.
\end{enumerate}
The state transitions between states are such as:
\begin{center}
  Event if conditions \trule actions.
\end{center}
They are triggered by an external event, e.g. the reception of a
message. The automaton is deterministic, i. e. the conditions of
triggering of outgoing transitions for the same state are mutually
exclusive. This behavioral design pattern allows to manage multi-party
dialogues. The proposals received by a coalition agent are processed
each in turn, even if they are stashed to be evaluated later on, but
their acceptance/rejection depend on the opinion of the other agents
which are concerned.

\begin{figure}
\begin{turn}{90}
    \scalebox{.6}{\input{figures/coalitionBehaviourSelectiveExact}}
\end{turn}
\caption{
Behaviour of the agent representing the coalition $\langle a, g \rangle$ 
in the selective mechanism without approximation. \mintinline{scala}{subgroups(g,min,max)} returns the subgroups of \mintinline{scala}{g} such as the size
is between  \mintinline{scala}{min} and \mintinline{scala}{max}.}
\label{algo:coalitionBehaviour}
\end{figure}
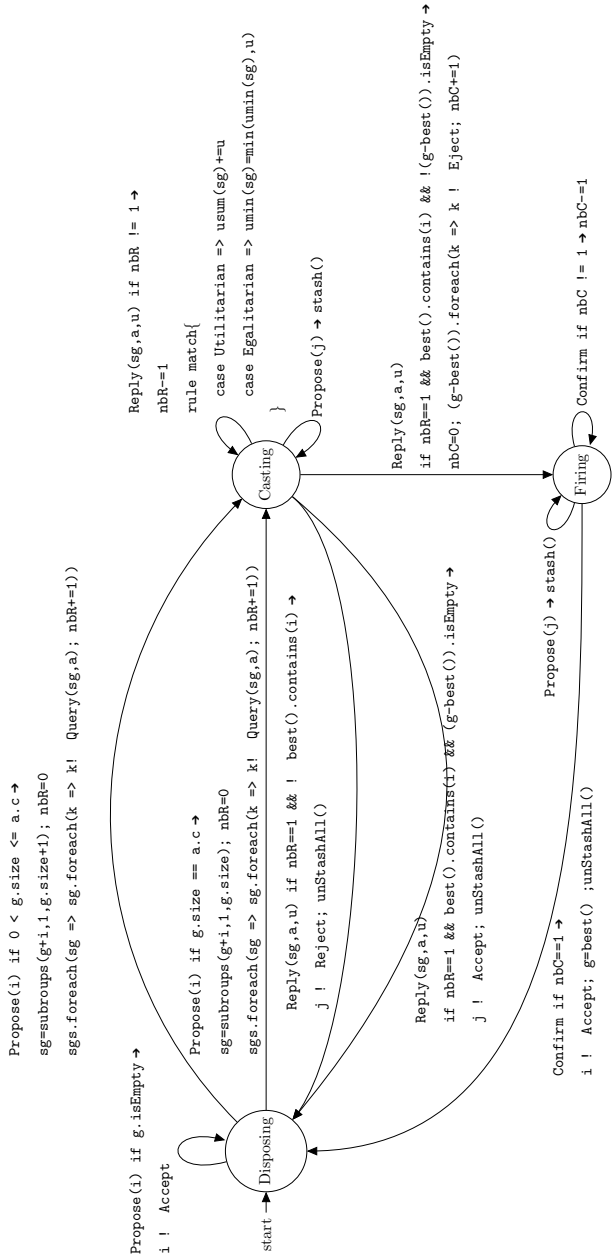

For instance, Fig.~\ref{algo:coalitionBehaviour} describes the
behaviour of the agent representing the coalition
$\langle a, g \rangle$ for the selective mechanism. According to this
behaviour, an proposal from $i$ which is received in the
\texttt{Disposing} state is accepted if the coalition is
empty. Otherwise, the agent evaluates this proposal in the
\texttt{Casting} by selecting the non-empty group according to the
capacity constraint and the objective of the coalition, i.e. the
social decision rule. Then, if the proposer is not a member of the
selected group, the proposal is rejected. Otherwise, the proposal is
accepted as soon as, if some members are excluded, the latter confirm
the notification of exclusion by the solver agent (in the
\texttt{Firing} state). In the other matching mechanisms, the
behaviour of coalition agents is similar with the exception of the
conditions and actions for the outgoing transitions of the
\texttt{Disposing} state. For instance, the assessment of the
coalition for the inclusive mechanism takes place only when the
capacity is reached.

\section{Experiments}
\label{sec:exp}

Our experiments\footnote{Our experiments have been performed with a 8
  core Intel (R) i7 2.8GHz and 16GB RAM.}  aim at evaluating: i) the
benefits of the structural properties of the problem for our
mechanisms; ii) the quality of the outcomes ; and iii) the speedup due
to their distribution.

Our prototype ScaIA (\url{https://github.com/maximemorge/ScaIA}) has
been implemented with the Scala progamming language and the Akka
toolkit (\url{http://akka.io}). The latter, which is based on the
actor model~\cite{hewitt73ijcai}, allows us to fill the gap between
the specification of our agent behaviours and their implementation.

Our mechanisms exploit the structure of the IA problem, which is,
contrary to the hedonic game, two-sided. In order to evaluate its
benefit, we translate an IA problem instance into:
\begin{enumerate}
\item a hedonic game by associating : i) one player for each activity
  such that all the sound coalitions are equally preferred; and ii)
  one player for each individual whose preferences are deduced from
  the preferences of the latter. Then, we compute a contractually
  individually stable (CIS) coalition;
\item a mixed integer quadratic programming (MIQP) problem
  (cf~\ref{sec:PL}) which is solved with
  IBM\textregistered~ILOG\textregistered~CPLEX\textregistered.
\end{enumerate}
We consider some IA problem instances with $2$ activities and
$m \in [2,20]$ individuals. For the sake of simplicity, all activities
have the same capacity ($c=\ceil{m/2}$).  We generate $100$
(pseudo-)random instances for each value of $m$.

First of all, we need to instanciate the definition of the affinity
function and the definition of the utility function. Inspired by our
practical application, we axiomatize the principle \textit{quo plus
  est eo hilarior, sed quo paucior eo melior cibus\footnote{In English
    \textit{the more the merrier, but the fewer the better fare}.}} by
computing the affinity of an individual $i$ for a group in the
following way:
\begin{align}
  w_i&: G(i) \rightarrow [-1,1] \nonumber\\
  \forall i \in  I ~ \forall g \in G(i)~  w_i(g) &= \frac{1}{m-1} \Sigma_{j \in g \setminus \{i\}} ~w_i(j)
\label{eq:wg}
\end{align}
This affinity, which is normalized ($w_i(g) \in [1.1]$), increases
with the size of the group if $i$ is attracted by all the members of
the group. This function, which evaluates the preferences of an
individual over $2^{m-1}$ groups, is additively separable and so its
representation is linear with respect to the number of individuals.
For the sake of simplicity, we define the utility function in the
following manner:
\begin{align}\label{eq:utility}
  u_i: G(i) \times A \cup \{\theta \} \rightarrow [-1,1] \nonumber \\
  \forall i \in I~ \forall g \in G(i) ~ \forall a \in A \cup \{\theta\},~u_i(g,a) = w_i(g) \oplus v_i(a)
\end{align}
Since $w_i(g), v_i(a) \in [-1,1]$, the utility is normalized
($u_i(g,a) \in [-1,1]$).

\begin{figure}
  \begin{center}
   \includegraphics[width=\columnwidth]{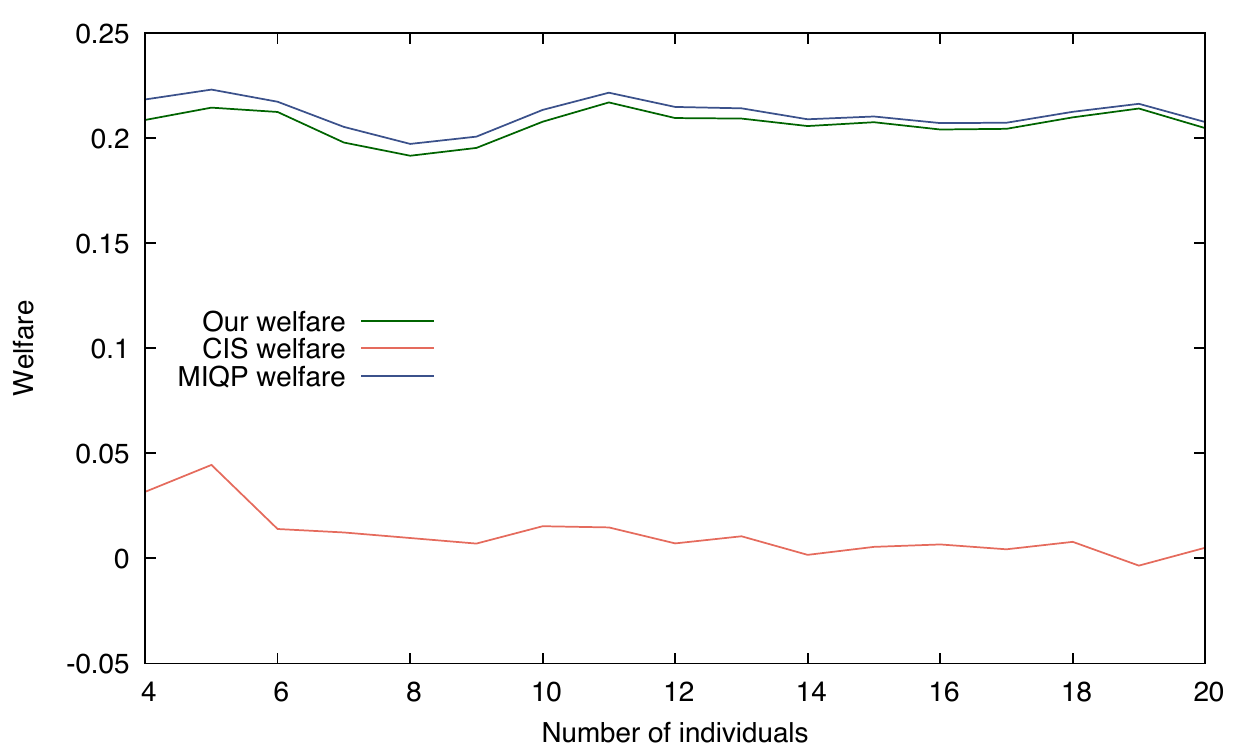}%
 \end{center}
 \caption{Utilitarian welfares of the selective
    approximation algorithm, of the MIQP solver and of the CIS solver.}%
  \label{exp:translationUtilitarianWelfare}
\end{figure}

\begin{figure}
  \begin{center}
   \includegraphics[width=\columnwidth]{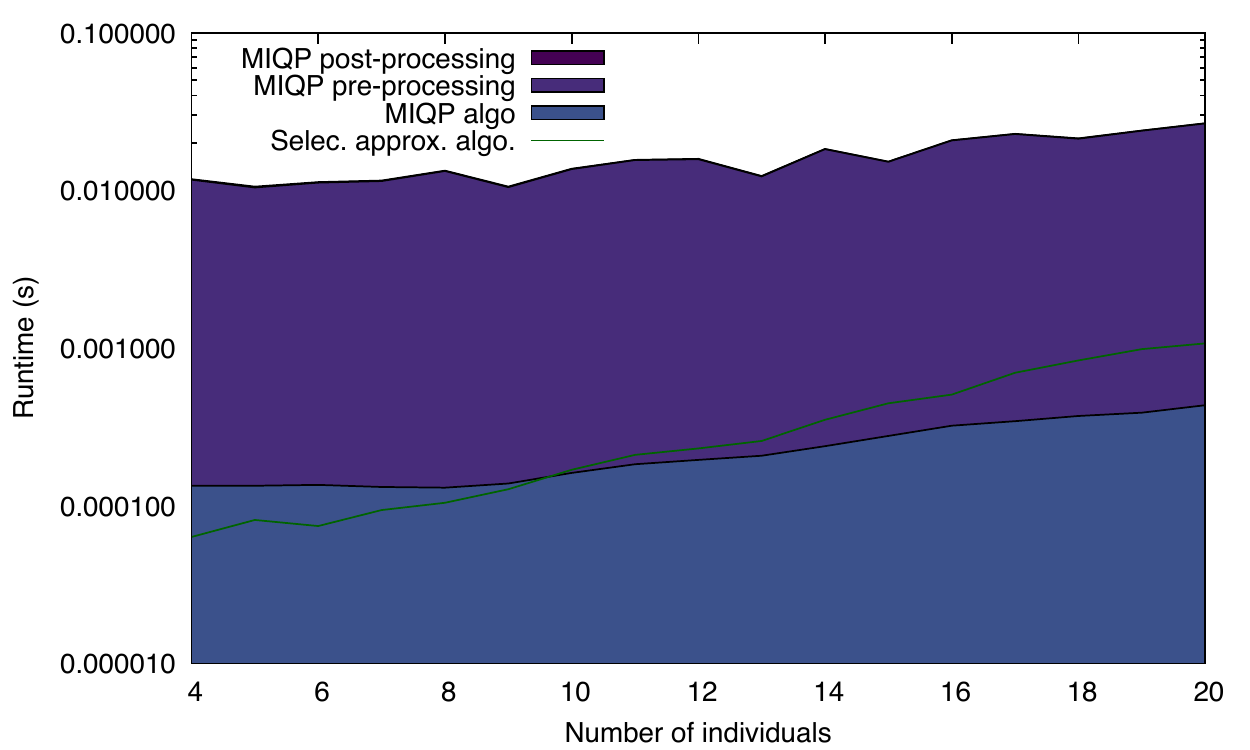}%
 \end{center}
 \caption{Runtimes of the selective algorithm with approximation and
   of the MIQP solver.}%
  \label{exp:translationUtilitarianMIQPTime}
\end{figure}

\begin{figure}
  \begin{center}
   \includegraphics[width=\columnwidth]{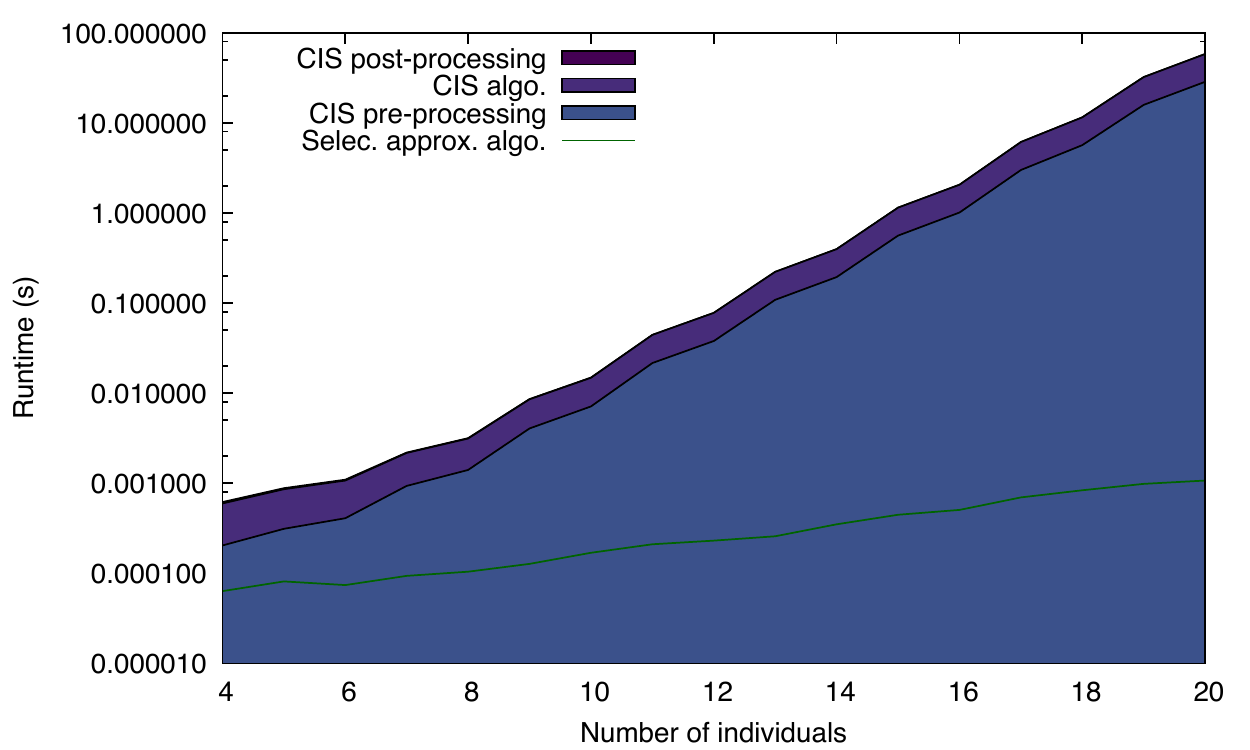}%
 \end{center}
 \caption{Runtimes of the selective algorithm with approximation and of the MIQP solver.}%
 \label{exp:translationTimeCISUtilitarian}
\end{figure}

Figure~\ref{exp:translationUtilitarianWelfare} compares the
utilitarian welfare reached by the selective algorithm with
approximation, the CIS algorithm and the MIQP
one. Figures~\ref{exp:translationUtilitarianMIQPTime} and
\ref{exp:translationTimeCISUtilitarian} compare the runtimes of these
algorithms with a logarithmic time scale and by distinguishing the
runtime for the translation of: i) the IA problem in a hedonic game or
in a mathematical program; and (ii) the solution.  The runtime of our
procedure benefits from the structural properties of the IA
problem. Moreover, the CIS algorithm is penalized by the generation of
RLC~\cite{balester04geb} which take exponential space. Even if the
selective mechanism does not always maximize the utilitarian welfare
computed by the MIQP, it is closed to. Moreover, the selective
mechanism is ($25$ times) faster (for $20$ individuals).

\begin{figure}
  \begin{center}
   \includegraphics[width=\columnwidth]{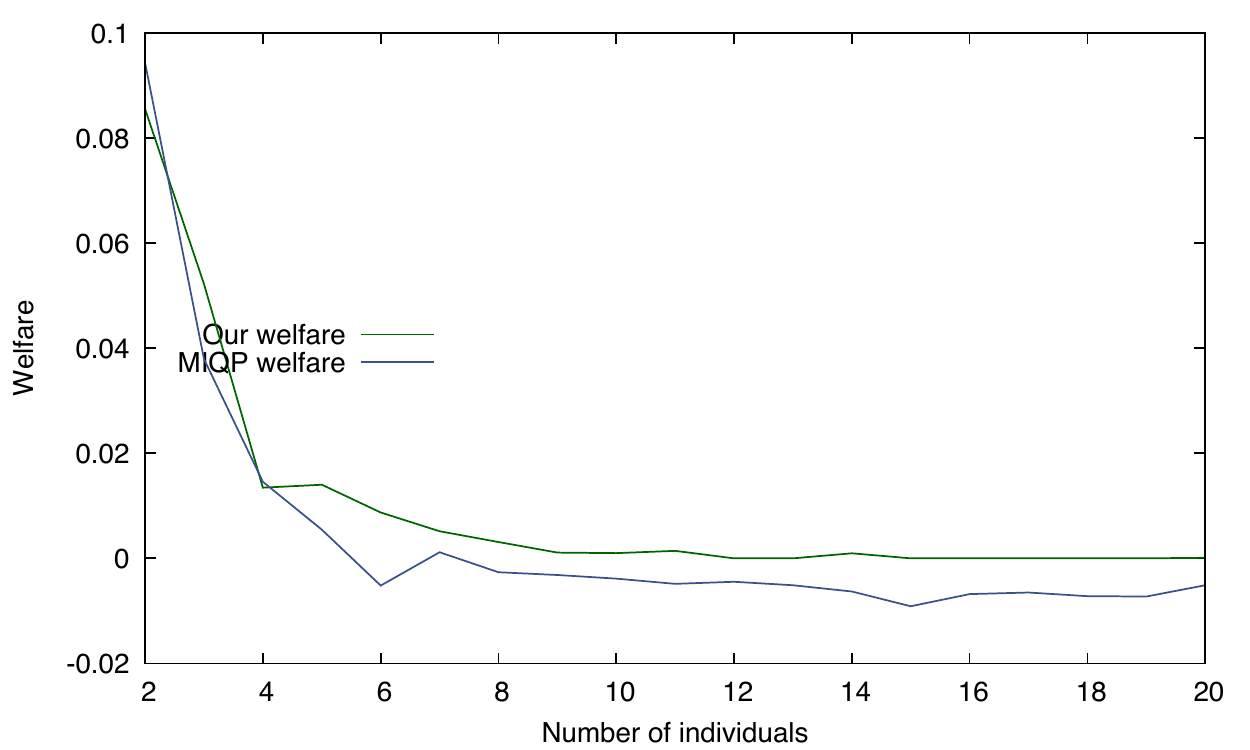}%
 \end{center}
 \caption{Egalitarian welfares of the inclusive algorithm and of the MIQP solver.}%
\label{exp:translationEgalitarianWelfare}
\end{figure}

\begin{figure}
  \begin{center}
    \includegraphics[width=\columnwidth]{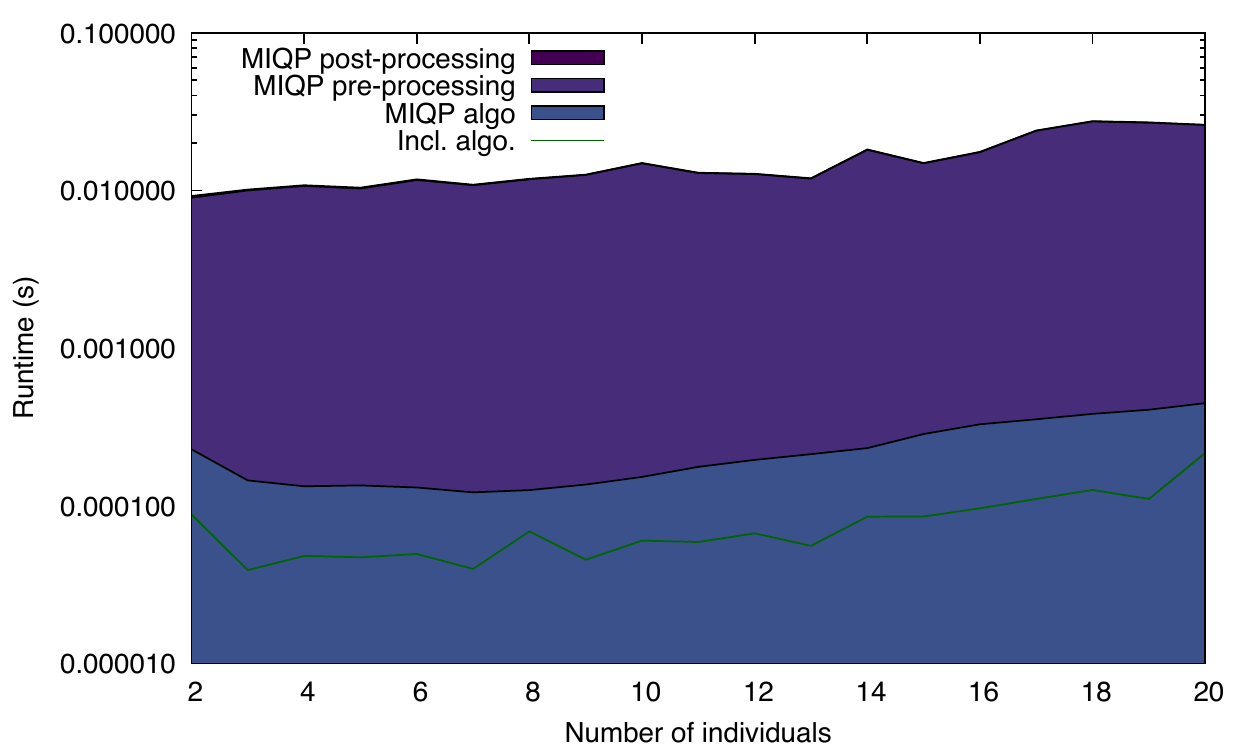}%
  \end{center}
  \caption{Runtime of the inclusive algorithm and of the MIQP
    solver.}%
 \label{exp:translationEgalitarianMIQPTime}
\end{figure}

Similarly, Figures~\ref{exp:translationEgalitarianWelfare}
and~\ref{exp:translationEgalitarianMIQPTime} compare the runtimes and
the egalitarian welfares reached by the inclusive algorithm and the
approximation algorithm (MIQP). The egalitarian welfare reached by our
algorithm is closed but better than the one approximated by MIQP while
being ($120$ times) faster (for $20$ individuals).

\begin{figure}
  \begin{center}
      \includegraphics[width=\columnwidth]{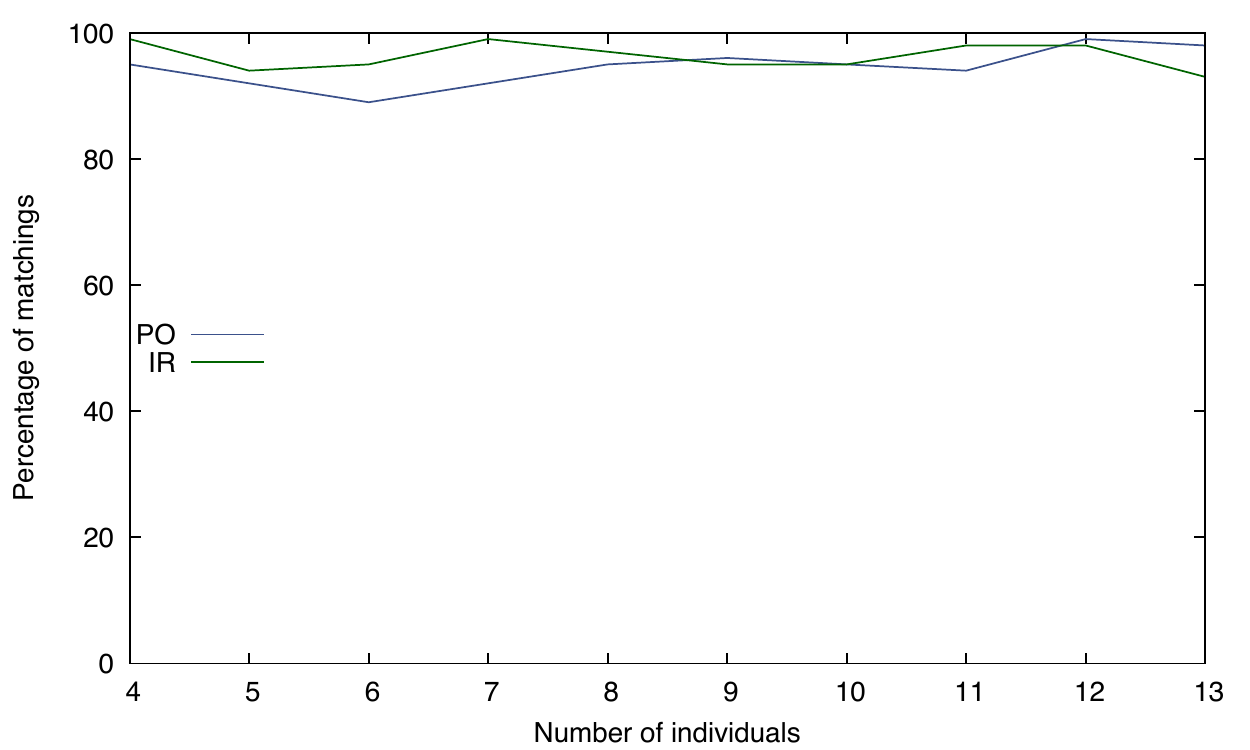}
  \end{center}
  \caption{Percentage of matchings computed by the selective algorithm
    with approximation which are Pareto-Optimal (PO) and individually
    rational (IR).}%
\label{exp:stability}
\end{figure}

Even if we do not have any theoretical results concerning the
stability of the outcome for the selective algorithm,
Figure~\ref{exp:stability} shows that $95 \%$ of the outcomes of the
selective algorithm with approximation are PO and $96 \%$ of them are
IR. We can note than we restrict ourselves to $n=2$ and $m \leq 13$
since the decision problem for the Pareto-optimality is not tractable.

Since the MIQP algorithms are not scalable, we have implemented a
local search algorithm~\cite{russell03ai} to be compared with our
algorithm. This hill-climbing algorithm, which starts with a sound and
random matching (where all the individuals are assigned to an
activity) and iteratively tries to improve the welfare. Two matchings
are neighbours if they are identical with an exception for one
individual which moves to another activity. If this new activity is
full, then all the swaps of individuals with the members of that
activity are considered.

We consider some IA problem instances with $n$ activities and $m$
individuals.  For the sake of simplicity, all the activities have the
same capacity ($c=\ceil{m/n}$).

Firstly, we focus on the utilitarian welfare of the outcome reached by
the selective mechanism with approximation.

\begin{figure}
  \begin{center}
      \includegraphics[width=\columnwidth]{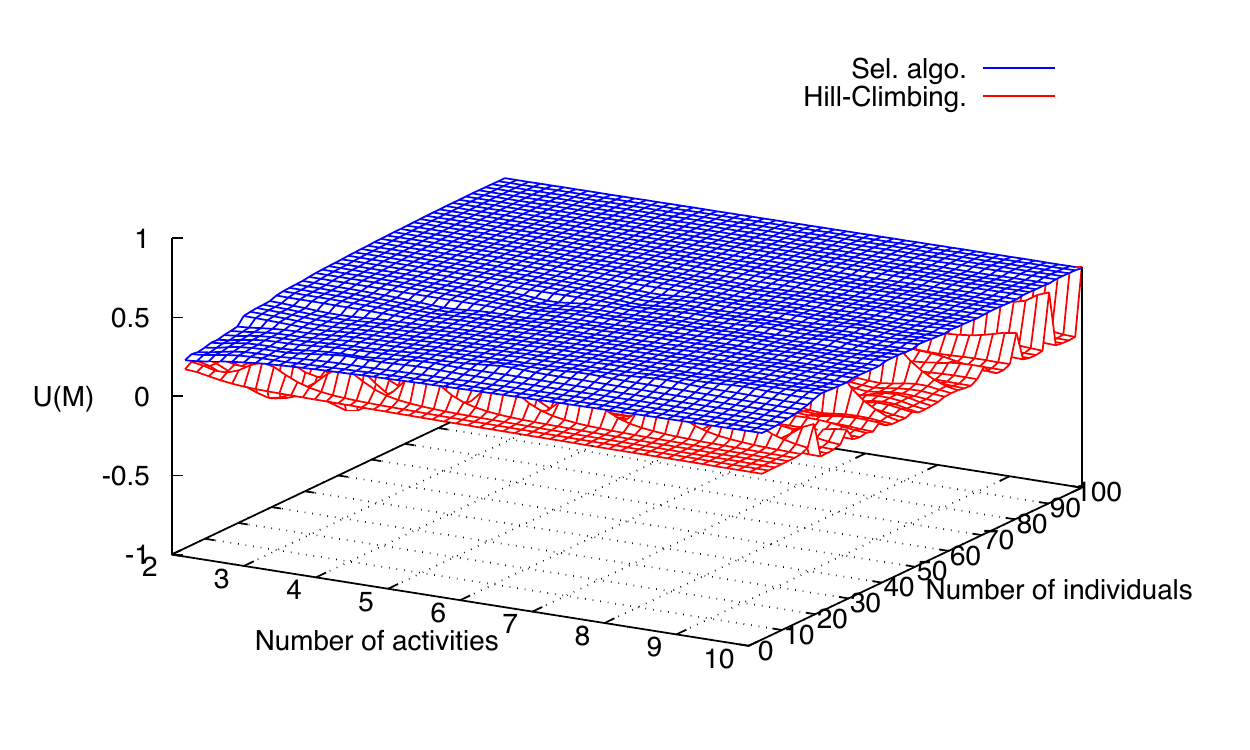}
  \end{center}
\caption{Utilitarian welfare for the selective/hill-climbing algorithms.}%
\label{exp:utilitarian}
\end{figure}

We can compare the utilitarian welfare of the matching reached by our
selective algorithm with the one reached by local
search. Fig.~\ref{exp:utilitarian} shows the average utilitarian
welfare of $100$ problem instances for each set of parameters (
$2\leq n\leq 10$ and $2\times n \leq m\leq 100$).  The selective
mechanism overreaches the local search. Indeed, the utilitarian
welfare for a IA problem is a function with many local optima.

\begin{figure}
  \begin{center}
      \includegraphics[width=\columnwidth]{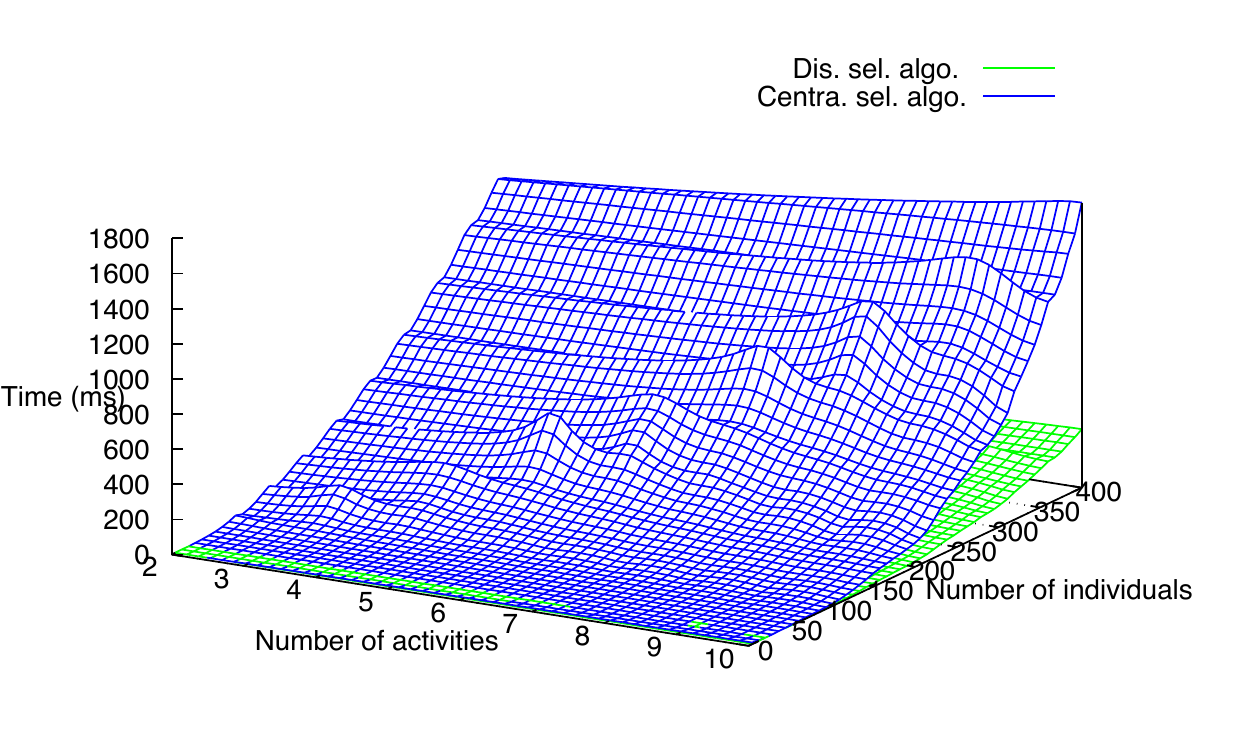}
  \end{center}
  \caption{Runtime for the centralized/decentralized selective
    algorithms with approximation.}%
\label{exp:timeUtilitarian}
\end{figure}

We can compare the runtime of the centralized version of our selective
algorithm with the decentralized one. Fig.~\ref{exp:timeUtilitarian}
shows the average runtime for each set of parameters (with
$2\leq n\leq 10$ and $2\times n\leq m\leq 400$).  While the
centralized algorithm is faster when the number of individuals is low
($\sim 20$), its runtime quickly grows with the number of individuals
($13$ ms for $100$ individuals and $10$ activities) while the runtime
of the distributed version is low ($4$ ms in the latter
case). Moreover, the runtime of the hill-climbing algorithm is too
high to be represented in Fig.~\ref{exp:timeUtilitarian} ($170,857$ ms
for $100$ individuals and $10$ activities). We can expect a higher
runtime if we adopt a local search method such as simulated annealing
without any warranty about the optimality of the outcome.

Secondly, we focus on the egalitarian welfare of the outcome reached
by the inclusive mechanism.

\begin{figure}
  \begin{center}
      \includegraphics[width=\columnwidth]{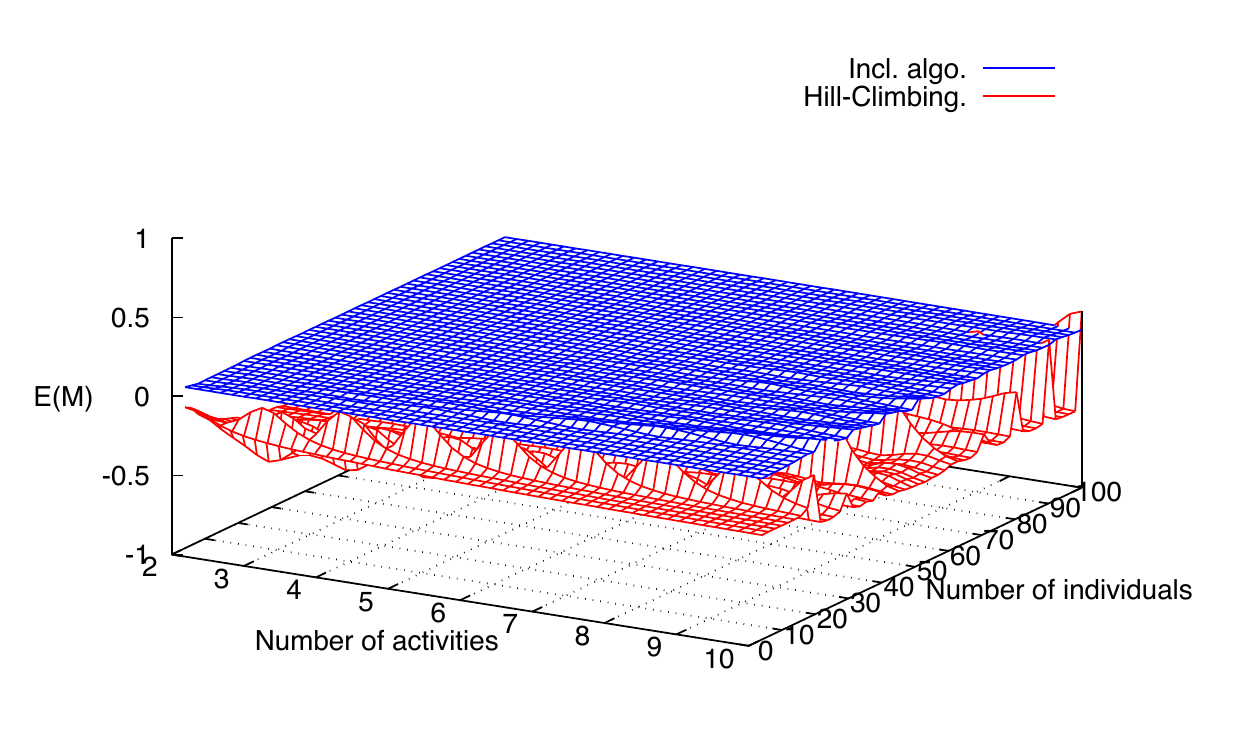}
  \end{center}
\caption{Egalitarian welfare for the inclusive/hill-climbing algorithms.}%
\label{exp:egalitarian}
\end{figure}

We can compare the egalitarian welfare of the matching reached by our
algorithm with the one reached by local
search. Fig.~\ref{exp:egalitarian} shows the average egalitarian
welfare of $100$ problem instances for each set of parameters (
$2\leq n\leq 10$ and $2\times n \leq m\leq 100$).  The inclusive
mechanism overreaches the local search. Indeed, the egalitarian
welfare for a IA problem is a function with many local optima.

\begin{figure}
  \begin{center}
      \includegraphics[width=\columnwidth]{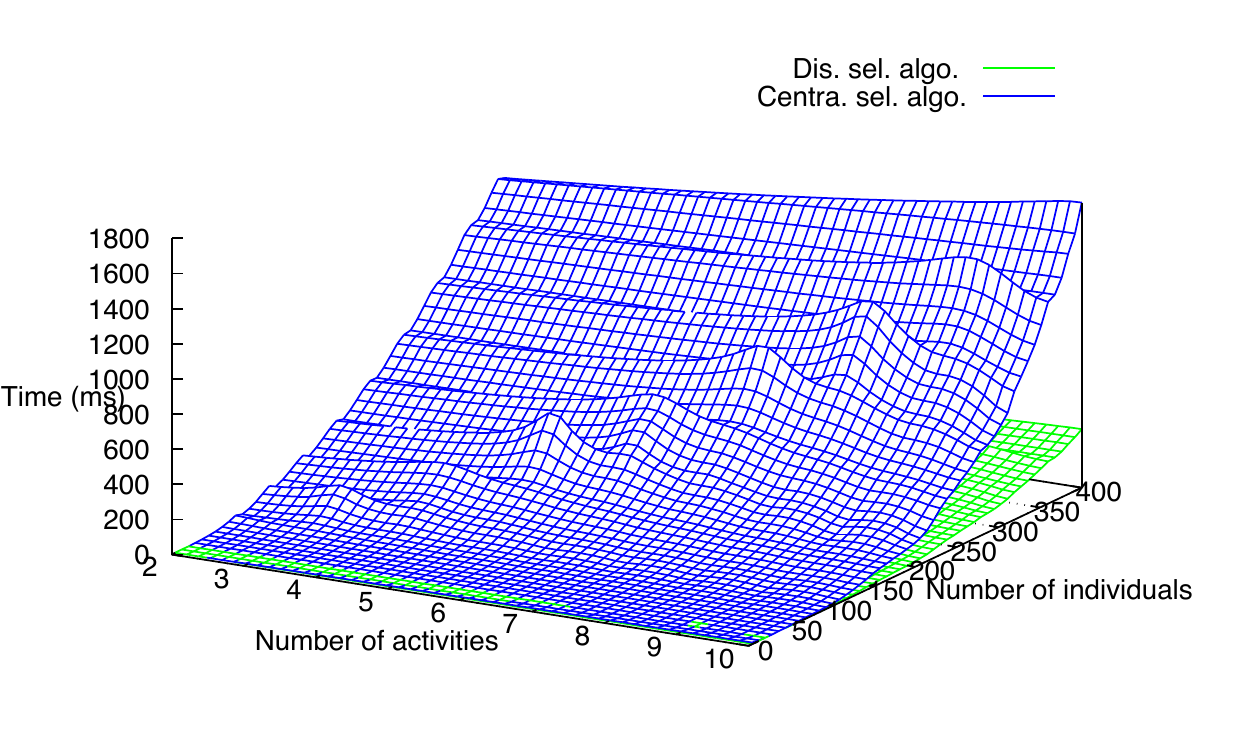}
  \end{center}
  \caption{Runtime for the centralized/decentralized inclusive algorithms.}%
\label{exp:timeEgalitarian}
\end{figure}

We can compare the runtime of the centralized version of our inclusive
algorithm with the decentralized one. Fig.~\ref{exp:timeEgalitarian}
shows the average runtime for each set of parameters (with
$2\leq n\leq 10$ and $2\times n\leq m\leq 400$).  While the
centralized algorithm is faster when the number of individuals is low
($\sim 40$), its runtime quickly grows with the number of individuals
($7$ ms for $100$ individuals and $10$ activities) while the runtime
of the distributed version is low ($2$ ms in the latter
case). Moreover, the runtime of the hill-climbing algorithm is too
high to be represented in Fig.~\ref{exp:timeEgalitarian} ($433,496$ ms
for $100$ individuals and $10$ activities). We can expect a higher
runtime if we adopt a local search method such as simulated annealing
without any warranty about the optimality of the outcome.

In summary, the outcomes reached by our algorithms seem to be closed
to the maximum utilitarian/egalitarian. Moreover, the distribution of
our algorithms allows to speedup its runtime.

\section{Practical application}
\label{sec:application}

Our practical application is concerned with a group of seniors who aim
at sharing some holidays. The objective is to maximize the
activities shared in order to improve the social cohesion and avoid
the isolation of seniors.

\begin{figure*}
  \begin{center}
      \includegraphics[angle=90,width=.7\textwidth]{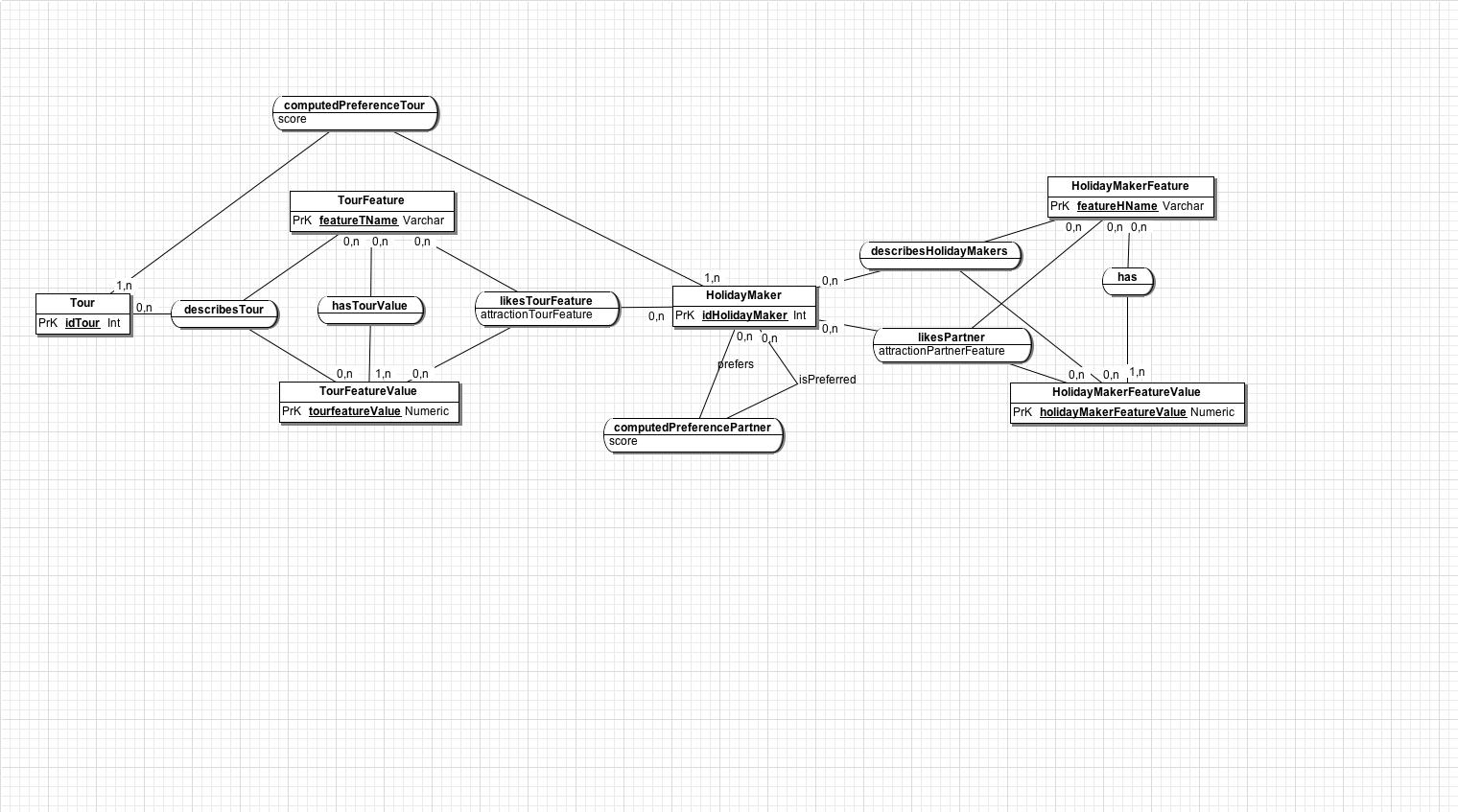}
    \end{center}
    \caption {Entity-relation model to capture the real-world data
      processed by our matching algorithms. Entities are represented
      by rectangles and relations by rectangles with rounded
      corners.}%
    \label{application:mcd}
\end{figure*}

According to the figure~\ref{application:mcd}, which represents the
entity-relation model of our database, each user is described by a set
of features: gender, age, hobbies, etc. Thus, each user indicates (or
not) the values which characterize him and he may inform its
attraction/repulsion towards these values for his peers. From this
knowledge we compute the affinities between users. In a way similar,
tours are described by feature values on which users express some
preferences. From this knowledge, we compute the interests of the
users for the tours.

Even if this scenario involves thousands of users, we have tested our
prototype with $102$ users and $10$
tours. Table~\ref{application:outcome} summarizes the runtimes and
welfares reach by all the matching algorithms, except the ones based
in mathematical optimization which are too slow. Whether we focus on
the utilitarian welfare or the egalitarian one, our algorithms are
faster and more efficient than hill-climbing. The runtimes of the
distributed versions of our algorithms are faster when the number of
individual integrations is sufficient.

\begin{table}
  \begin{center}
    \begin{tabular}{c|r|r}
      Algorithm & Welfare & Runtime (ms)\\
      \hline
      Selec. with appr. & $U(M)=0.154$ & $263$ \\
      Dis. selec. with appr. & $U(M)=0.156$ & $210$ \\
      Hill-climbing & $U(M)=-0.066$ & $49~116$ \\
      \hline
      Inclusive & $E(M)=0.0$ & $22$ \\
      Dis. inclusive & $E(M)=0.0$ & $44$ \\      
      Hill-climbing & $E(M)=-0.433$ & $20~643$\\
    \end{tabular}
  \end{center}
  \caption{Welfares and runtimes of the matching algorithms with the
    real-world data.}%
  \label{application:outcome}
\end{table}

\section{Discussion}
\label{sec:discussion}

We have presented here a generic problem where some individuals must
be assigned to the activities they enjoy with their favorite
partners. Even if stability is desirable, it is not guaranteed. By
contrast, Pareto-optimality seems not to be discriminative. This is
the reason why we evaluated a matching from the collective viewpoint
using concepts derived from social choice theory. Others concepts may
be relevant, whether they come from Economy (envy-freeness,
strategy-proofness, etc.) or inspired by social psychology such as the
concept of social cohesion sketched here.

Even if a worst-case study is required here, our experiments show that
social welfares are functions with many local optima. That is the
reason why we have proposed here two heuristics, one for each of the
notions of social welfare. They returns sound matchings in accordance
with a social decision rule (either utilitarian or egalitarian) for
the first and which is socially cohesive for the second.  These
mechanisms are based on the organizational aspect of agent-based
modelling by explicitly introducing an intermediate level between the
individual and the society, i.e. a coalitions of individuals who share
a common interest for an activity. In accordance with the principle of
subsidiarity, the assessment of a coalition is performed by an agent
representing the objectives and constraints of a group of individuals.

Our empirical results, in particular with real-world data, illustrate
that multiagent resolution methods can be more effective than general
optimization methods which require problem reformulation. Beyond a
simple decentralization that, as here, speedups its runtime, the
modeling of agents behaviours allows to introduce a variety of
strategies for implementing the constraints and the objectives of the
different agent coalitions.

In future works, we plan to take into account the relational aspect of
the agent-based modelling to distribute the negotiations between
neighbours in the social network.

\section*{References}
\bibliography{main}

\appendix

\section{Linear programming}
\label{sec:PL}

Let us consider a IA problem instance of size $(m,n)$.

In order to find a maximum utilitarian matching, we can consider
quadratic programming, i. e. a method of optimization with a
mathematical model that can be represented by:
\begin{enumerate}
\item $n \times m$ decision variables $x_{ia} \in \{0,1\}$ such that,
  \begin{align}
    x_{ia} =&\begin{cases}
      1 & \text{ if $i$ is assigned to $a$},\\
      0 & \text{otherwise ;}
    \end{cases}
  \end{align}
\item $m$ constraints representing the mutual exclusion of
  individual assignements,
  \begin{equation}
    \Sigma_{a \in A} x_{ia} \leq 1 ;\label{eq:exclusion}
  \end{equation}
\item $n$ constraints for the soundness of the matching,
  \begin{equation}
    \label{eq:soundness}
    \Sigma_{i \in I} x_{ia} \leq c_a.
  \end{equation}
\item the objective function to maximize:
  \begin{equation}
    \frac{1}{m} \sum_{i \in I} \sum_{a \in A} \bigg[  \frac{1}{2}  \Big( x_{ia} v_i(a)  + \frac{1}{m-1} \sum_{j \neq i} x_{ja} w_{i}(j)  \Big) \bigg] \label{eq:uswobj}
  \end{equation}
\end{enumerate}

In order to find a maximum egalitarian matching, we can consider
the following linear program:
\begin{enumerate}
\item $n \times m$ decision variables $x_{ia} \in [-1,1]$ such that,
  \begin{align}
    x_{ia} =&\begin{cases}
      1 & \text{ if $i$ is assigned to $a$},\\
      0 & \text{otherwise ;}
    \end{cases}
  \end{align}
\item a decision variable $uMin \in [-1;1]$ representing the minimal
  utility of individuals;
\item $m$ constraints representing the mutual exclusion of
  individual assignements,
  \begin{equation}
    \Sigma_{a \in A} x_{ia} \leq 1 ;\label{eq:exclusion}
  \end{equation}
\item $n$ constraints for the soundness of the matching,
  \begin{equation}
    \label{eq:soundness}
    \Sigma_{i \in I} x_{ia} \leq c_a.
  \end{equation}
\item $m$ constraints for the minimal utilities of the individuals:
  \begin{equation}
    \sum_{a \in A} \bigg[  \frac{1}{2}  \Big( x_{ia} v_i(a)  + \frac{1}{m-1} \sum_{j \neq i} x_{ja} w_{i}(j)  \Big) \bigg] \geq uMin\label{eq:uMin}
  \end{equation}
\item the objective function to minimize: $uMin$.
\end{enumerate}
This program allows to
implement an approximation algorithm.

\end{document}

%% file: figures/coalitionBehaviourSelectiveExact.tex
\begin{tikzpicture}[>=triangle 45]

  \node (Disposing) at (0,7) [state,initial] {Disposing}; 

  \node (Casting) at (15,7) [state] {Casting}; 

  \node (Firing) at (15,0) [state] {Firing}; 

  \path[->] 
  (Disposing) 
  edge [loop above] node[align=left] 
  {\mintinline{scala}{Propose(i) if g.isEmpty} 
    \trule\\ 
    \mintinline{scala}{i ! Accept}
  } 
  () 

  edge [above,out=45,in=135] node[align=left] 
  {\mintinline{scala}{Propose(i) if 0 < g.size <= a.c} 
    \trule\\ 
    \mintinline{scala}{sg=subroups(g+i,1,g.size+1); nbR=0}\\
    \mintinline{scala}{sgs.foreach(sg => sg.foreach(k => k! Query(sg,a); nbR+=1))}\\
  } 
  (Casting) 
    
  edge [above,out=0,in=180, looseness=0] node[align=left] 
  {\mintinline{scala}{Propose(i) if g.size == a.c} 
    \trule\\ 
    \mintinline{scala}{sg=subroups(g+i,1,g.size); nbR=0}\\
    \mintinline{scala}{sgs.foreach(sg => sg.foreach(k => k! Query(sg,a); nbR+=1))}
  } 
  (Casting) 

  (Casting) edge [right, out=330, in=300,looseness=8] node 
  {\mintinline{scala}{Propose(j)}
    \trule 
    \mintinline{scala}{stash()}} 
  (Casting) 
  
  edge [right, out=25, in=55, looseness=8] node[align=left] 
  {\mintinline{scala}{Reply(sg,a,u) if nbR != 1}
    \trule\\
    \mintinline{scala}{nbR-=1}\\
    \mintinline{scala}{rule match}\{\\
    ~~~\mintinline{scala}{case Utilitarian => usum(sg)+=u}\\
    ~~~\mintinline{scala}{case Egalitarian => umin(sg)=min(umin(sg),u)}\\
    \}\\
  } (Casting) 
  
  edge [above,out=-130,in=-40, looseness=.5] node[align=left, yshift=.5cm] 
  {\mintinline{scala}{Reply(sg,a,u) if nbR==1 \&\& ! best().contains(i)}
    \trule\\ 
    \mintinline{scala}{j ! Reject; unStashAll()}} 
  (Disposing) 

  edge [above,out=-130,in=-40, looseness=1.2] node[align=left, yshift=-1cm] 
  {\mintinline{scala}{Reply(sg,a,u)}\\
    \mintinline{scala}{if nbR==1 \&\& best().contains(i) \&\& (g-best()).isEmpty} 
    \trule\\ 
    \mintinline{scala}{j ! Accept; unStashAll() }}
  (Disposing) 

  edge [right] node[align=left] 
  {\mintinline{scala}{Reply(sg,a,u)}\\
    \mintinline{scala}{if nbR==1 \&\& best().contains(i) \&\& !(g-best()).isEmpty}
    \trule\\ 
    \mintinline{scala}{nbC=0; (g-best()).foreach(k => k ! Eject; nbC+=1)}}
  (Firing) 

  (Firing) 
  edge [left, out=165, in=135,looseness=8] node 
  {\mintinline{scala}{Propose(j)}
    \trule 
    \mintinline{scala}{stash()}} 
  (Firing) 

  edge [loop right] node[align=left] 
  {\mintinline{scala}{Confirm if nbC != 1} 
    \trule 
    \mintinline{scala}{nbC-=1}
  } 
  () 

  edge [below, out=-180,in=-90,looseness=1] node[align=left] 
  {\mintinline{scala}{Confirm if nbC==1}
    \trule\\
    \mintinline{scala}{i ! Accept; g=best() ;unStashAll()}} 
  (Disposing) 

;
\end{tikzpicture}

%% file: main.bbl
\begin{thebibliography}{10}
\expandafter\ifx\csname url\endcsname\relax
  \def\url#1{\texttt{#1}}\fi
\expandafter\ifx\csname urlprefix\endcsname\relax\def\urlprefix{URL }\fi
\expandafter\ifx\csname href\endcsname\relax
  \def\href#1#2{#2} \def\path#1{#1}\fi

\bibitem{drerze80econometrica}
J.~Dreze, J.~Greenberg, Hedonic coalitions: Optimality and stability,
  Econometrica 48 (1980) 987--1003.

\bibitem{aziz13ai}
H.~Aziz, F.~Brandt, H.~G. Seedig, Computing desirable partitions in additively
  separable hedonic games, Artificial Intelligence Journal 195 (2013) 316--334.

\bibitem{balester04geb}
C.~Ballester, {{NP}}-completeness in hedonic games, Games and Economic Behavior
  49~(1) (2004) 1--30.

\bibitem{schelling80strategy}
T.~C. Schelling, The strategy of conflict, Harvard university press, 1980.

\bibitem{darmann12wine}
A.~Darmann, E.~Elkind, S.~Kurz, J.~Lang, J.~Schauer, G.~Woeginger, Group
  activity selection problem, in: Proceedings of the 8th International
  Conference on Internet and Network Economics, Springer Berlin Heidelberg,
  Liverpool, UK, 2012, pp. 156--169.

\bibitem{igarashi17aaai}
A.~Igarashi, D.~Peters, E.~Elkind, Group activity selection on social networks,
  in: Proc. of 31th {{AAAI}} Conference on Artificial Intelligence, San
  Francisco, California USA, 2017, pp. 565--571.

\bibitem{gale62ams}
D.~Gale, L.~S. Shapley, College admissions and the stability of marriage, The
  American Mathematical Monthly 69 (1962) 9--14.

\bibitem{manlove14book}
D.~F. Manlove, Algorithmics of Matching Under Preferences, World Scientific,
  2014.

\bibitem{everaere13aamas}
P.~Everaere, M.~Morge, G.~Picard, {Minimal Concession Strategy for Reaching
  Fair, Optimal and Stable Marriages}, in: Proc. of AAMAS, 2013, pp.
  1319--1320.

\bibitem{nongaillard16ecai}
A.~Nongaillard, S.~Picault, Multilevel agent-based modelling for assignment or
  matching problems, in: Proc. of ECAI, 2016, pp. 1561--1562.

\bibitem{boutilier05ai}
C.~Boutilier, I.~Caragiannis, S.~Haber, T.~Lu, A.~D. Procaccia, O.~Sheffet,
  Optimal social choice functions: A utilitarian view, Artificial Intelligence
  Journal 227 (2015) 190--213.

\bibitem{sen70socialchoice}
A.~K. Sen, Collective Choice and Social Welfare, North-Holland, 1970.

\bibitem{moulin02fair}
H.~Moulin, Fair Division and Collective Welfare, The MIT Press, 2002.

\bibitem{zaccaro88jsp}
S.~J. Zaccaro, C.~A. Lowe, Cohesiveness and performance on an additive task:
  Evidence for multidimensionality, The Journal of Social Psychology 128~(4)
  (1988) 547--558.
\newblock \href {http://dx.doi.org/10.1080/00224545.1988.9713774}
  {\path{doi:10.1080/00224545.1988.9713774}}.

\bibitem{clinger81foundations}
W.~D. Clinger, Foundations of actor semantics, Ph.D. thesis, Massachusetts
  Institute of Technology (1981).

\bibitem{hewitt73ijcai}
C.~Hewitt, P.~Bishop, R.~Steiger, A universal modular {{ACTOR}} formalism for
  artificial intelligence, in: Proc. of the 3rd International Joint Conference
  on Artificial Intelligence, Morgan Kaufmann Publishers Inc., San Francisco,
  CA, USA, 1973, pp. 235--245.

\bibitem{russell03ai}
S.~Russell, P.~Norvig, Artificial Intelligence: A Modern Approach, Pearson
  Education, 2003, Ch. Chapter 4: Informed Search and Exploration, pp. 94--136,
  2nd edition.

\end{thebibliography}
